# Gas assisted binary black hole formation in AGN discs

Henry Whitehead ®,[1]★ Connar Rowan ®,[2] Tjarda Boekholt ®[2,3] and Bence Kocsis ®[2,4]

[1]*Department of Physics, Astrophysics, University of Oxford, Denys Wilkinson Building, Keble Road, Oxford OX1 3RH, UK*
[2]*Rudolf Peierls Centre for Theoretical Physics, Clarendon Laboratory, University of Oxford, Parks Road, Oxford OX1 3PU, UK*
[3]*NASA Ames Research Center, Moffett Field, CA 94035, USA*
[4]*St Hugh's College, St Margaret's Rd, Oxford OX2 6LE, UK*

## ABSTRACT

We investigate close encounters by stellar mass black holes (BHs) in the gaseous discs of active galactic nuclei (AGNs) as a potential formation channel of binary black holes (BBHs). We perform a series of 2D isothermal viscous hydrodynamical simulations within a shearing box prescription using the Eulerian grid code Athena++. We co-evolve the embedded BHs with the gas keeping track of the energetic dissipation and torquing of the BBH by gas gravitation and inertial forces. To probe the dependence of capture on the initial conditions, we discuss a suite of 345 simulations spanning BBH impact parameter ($b$) and local AGN disc density ($\rho_0$). We identify a clear region in $b - \rho_0$ space where gas assisted BBH capture is efficient. We find that the presence of gas leads to strong energetic dissipation during close encounters between unbound BHs, forming stably bound eccentric BBHs. We find that the gas dissipation during close encounters increases for systems with increased disc density and deeper periapsis passages $r_p$, fitting a power law such that $\Delta E \propto \rho_0^\alpha r_p^\beta$, where $\{\alpha, \beta\} = \{1.01 \pm 0.04, -0.43 \pm 0.03\}$. Alternatively, the gas dissipation is approximately $\Delta E = 4.3 M_d v_H v_p$, where $M_d$ is the mass of a single BH minidisc just prior to the encounter when the binary separation is $2r_H$ (two binary Hill radii), $v_H$ and $v_p$ are the relative BH velocities at $2r_H$ and at the first closest approach, respectively. We derive a prescription for capture which can be used in semi-analytical models of AGN. We do not find the dissipative dynamics observed in these systems to be in agreement with the simple gas dynamical friction models often used in the literature.

## 1 INTRODUCTION

Mergers of binary black holes (BBHs) have been detected by ground based observatories LIGO/Virgo/KAGRA by the gravitational waves (GWs) they emit (Abbott et al. 2016, 2019, 2020a, b, c, 2022). Analysis of these signals reveals a spectrum of black hole (BH) masses that are in tension with stellar evolution models, which predict a limit to BH mass as supernova remnants due to pair-instability (Rakavy & Shaviv 1967). While continual accretion after the BH's initial formation may be sufficient to reach these masses (Safarzadeh & Haiman 2020), an alternative origin for these 'mass gap' BHs has been hypothesized to be mergers between BHs (O'Leary et al. 2006; O'Leary, Meiron & Kocsis 2016; Gerosa & Berti 2019). Such mergers are expected to occur in dense stellar clusters due to many-body encounters (Mouri & Taniguchi 2002). Outside of stellar clusters, an additional channel for BH merger lies within the gaseous discs surrounding active galactic nuclei (AGNs) (McKernan et al. 2012, 2014; Stone, Metzger & Haiman 2017; Secunda et al. 2019; Yang et al. 2019; Gröbner et al. 2020; McKernan et al. 2020; Tagawa, Haiman & Kocsis 2020a; Tagawa et al. 2020b, 2021a, b). In this region we expect a significant number density of BHs within ∼ 1 pc (Bahcall & Wolf 1976; Miralda-Escudé & Gould

2000). BHs on orbits inclined to the disc are predicted to be driven into the disc plane by gas dynamical friction, further increasing the likelihood of BH–BH interactions (Bartos et al. 2017; Panamarev et al. 2018). Gas torques on BHs embedded in the discs may result in overdensities of BHs on similar orbits in regions named migration traps (McKernan et al. 2012; Bellovary et al. 2016). BH interactions in AGN are of specific interest due to the potential for detectable perturbations to the GW signal (Zwick et al. 2024) and accompanying electromagnetic emission (McKernan et al. 2019; Kimura, Murase & Bartos 2021; Tagawa et al. 2023a; Tagawa, Kimura & Haiman 2023b; Rodríguez-Ramírez et al. 2024; Tagawa et al. 2024).

Binary capture can occur when a sufficient proportion of the binary energy is dissipated during a close encounter. In planetary/stellar systems this can occur due to tidal deformation of the interacting bodies (Fabian, Pringle & Rees 1975). In the case of BHs, if the encounter is suitably close, GW emission can drive dissipation (Hansen 1972). The requirement for an exceptionally close encounter to drive significant GW dissipation results in a very small capture cross-section to perturbative effects alone (O'Leary, Kocsis & Loeb 2009), but the likelihood of such a capture occurring is expected to be enhanced during Jacobi interactions where multiple close encounters between BHs can increase the total dissipation (Boekholt, Rowan & Kocsis 2023a). The presence of gas during a close encounter provides an avenue for binary formation by exerting a drag force through gravitation (gas dynamical friction; Ostriker 1999) or via accretion of gas momentum. In this paper, we continue to examine this process using high resolution hydrodynamics simulations.

★ E-mail: henry.whitehead@physics.ox.ac.uk

Previous studies into the formation and evolution of BBHs span a variety of methodologies. Even for studies of BBHs in isolation, the long-term behaviour varies depending on the initial conditions and physics included within. Studies of thicker circumbinary discs with a scale height ratio $H/R = 0.1$ tend to conclude that binaries should outspiral due to net positive torques (Tang, MacFadyen & Haiman 2017; Moody, Shi & Stone 2019; Muñoz, Miranda & Lai 2019; Tiede et al. 2020)). This does not seem to hold for studies of cooler, thinner discs (Heath & Nixon 2020; Tiede et al. 2020), which predict negative net torques. Other simulation specific uncertainties such as choice of resolution, sink rate (Tang et al. 2017) or softening lengths (Li et al. 2021) also seem to affect the binary evolution. Conclusions drawn by more recent studies seem to have converged towards positive torques softening gaseous BBHs over time, at least for circular equal-mass systems (Duffell et al. 2020; Muñoz & Lithwick 2020; Westernacher-Schneider et al. 2022)). Westernacher-Schneider et al. (2024) suggests that SMBH binaries may possess an electromagnetic signature that could differentiate them from single SMBH quasi-periodic sources due to the eccentric minidiscs formed in the binary case.

The situation is complicated further for studies that attempt to analyse the behaviour of BBHs embedded within the shearing flow of an AGN disc. Early numerical simulations suggest that the binary should be hardened by the AGN disc gas during its inward migration (Baruteau, Cuadra & Lin 2011). Later studies (Li et al. 2021) found that the softening length can affect whether or not a binary hardens, showing prograde binaries would harden only if the circumsingle discs were insufficiently resolved. A trio of papers featuring 2D shearing box studies of embedded BBH in AGN found hardening by gas for a range of thermal and viscous properties (Li & Lai 2022, 2023a, b). More recent works have also considered the effect of inclination in the inner binary (Dittmann, Dempsey & Li 2023), finding inclined binaries are driven to alignment with the AGN disc over time.

More pertinent to this study are investigations into the formation of the BBHs themselves. Numerical methods for these formation events have considered dynamical friction models (DeLaurentiis, Epstein-Martin & Haiman 2023; Qian, Li & Lai 2023; Rozner, Generozov & Perets 2023), smoothed-particle hydrodynamics (Rowan et al. 2023b) or Eulerian grid codes (Li et al. 2023). Rowan et al. (2023b, henceforth CR1) includes study of the subsequent binary evolution and reaches agreement with pre-formed embedded binary studies that hardening is dependent on orbital rotation (prograde/retrograde). CR1 reveals the mechanism for gaseous capture, namely the formations of minidiscs around BHs, the importance of the first encounter and how the interplay between the gravitational gas dissipation and accretion determines the outcome. Rowan et al. (2023a, henceforth CR2) continues this work to consider how the inclusion of gas affects the conclusions of Boekholt et al. (2023a), by probing the interplay between the SMBH effects and gas gravitation/accretion over a wider range of initial conditions.

In this paper, we study the interactions between two initially isolated BHs embedded within an AGN disc using a grid of localized 2D viscous hydrodynamics simulations. We examine how the local density of the AGN disc and the initial radial separation of the BHs affects the likelihood of capture. We perform a total of 345 simulations spanning these two parameters. Compared to previous studies, we provide a finer sampling of parameter space by an order of magnitude, while also testing the capture scenario with a high resolution grid-based hydro method where the gas flow is better resolved. We first describe the computational methodology in Section 2, followed by detailing the system's initial conditions in Section 3. We present our fiducial model in Section 4, discussing the general chronology of a capture as well as the energetic and angular momentum evolution. We follow this with a broader discussion of simulations with varying initial conditions in Section 5. In Section 6 we describe a novel methodology for predicting the outcome of binary capture without the need for full hydrodynamical simulation, as derived from our numerical models. We discuss the general findings of the investigation in Section 7 and summarize our conclusions in Section 8. Brief asides to specific routines can be found in the Appendices A–G.

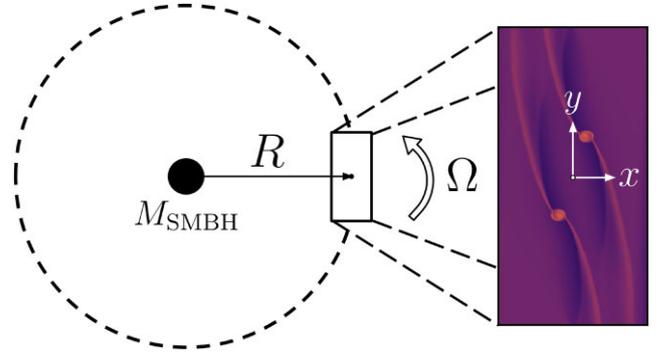

**Figure 1.** Schematic description of the shearing box prescription. Only a Cartesian cut-out of the disc is simulated, in a frame that rotates around the SMBH at the Keplerian rate $\Omega^2 = GM_{SMBH}/R^3$. Pictured within the shearing frame is a generic hydrodynamic snapshot where gas density is displayed with a logarithmic colour map, displaying two BHs approaching each other within their respective minidiscs.

## 2 COMPUTATIONAL METHODS

We use the Eulerian GRMHD code `Athena++` (Stone et al. 2020) to perform our hydrodynamical simulations. Similar to previous studies (e.g. CR1), we neglect any effects associated with gas self-gravity, relativity, radiation, and magnetism, restricting ourselves to an isothermal equation of state for simplicity. We utilize a second-order van Leer predictor–corrector integrator with a piecewise linear method (PLM) spatial reconstruction and Roe's linearized Riemann solver. A more detailed description of `Athena++`'s integration, reconstruction and solving routines can be found in Stone et al. (2020). Our simulation tracks a 2D rectangular patch of disc (shearing box) co-rotating at the Keplerian angular frequency $\Omega$ with respect to the SMBH. The geometry and orientation of this patch are detailed in Fig. 1. Within this box our BHs function as Newtonian point masses which are self-consistently propagated through the flow. The natural length scales in the shearing box are the Hill radii of a single BH or BBH ($r_{H, s}$ and $r_H$ respectively)

$$r_{H,s} = R\left(\frac{m_{BH}}{3M_{SMBH}}\right)^{\frac{1}{3}}, \quad r_H = R\left(\frac{M_{bin}}{3M_{SMBH}}\right)^{\frac{1}{3}} \quad (1)$$

where $m_{BH}$ and $M_{bin}$ are the masses of a single BH or a BBH orbiting a central SMBH of mass $M_{SMBH}$ at a radius $R$.

### 2.1 Gas dynamics

Due to our rotating frame of reference the gas in the simulation feels, along with the standard hydrodynamical forces, contributions from gravitation towards the stellar mass black holes and inertial forces from the uniform rotation of the shearing box. Evolving the fluid therefore requires solving the extended Navier–Stokes equations:

$$\frac{\partial \rho}{\partial t} + \nabla \cdot (\rho \boldsymbol{u}) = 0 \tag{2}$$

$$\frac{\partial (\rho \boldsymbol{u})}{\partial t} + \nabla \cdot (\rho \boldsymbol{u}\boldsymbol{u} + \boldsymbol{P}\boldsymbol{I} + \boldsymbol{\Pi}) = \rho \left( \boldsymbol{a}_{\text{SMBH}} - \nabla \phi_{\text{BH}} \right), \tag{3}$$

where we have introduced $\rho$, $P$, $\boldsymbol{u}$, and $\boldsymbol{\Pi}$ as the gas density, pressure, velocity, and viscous stress tensor

$$\Pi_{ij} = \rho \nu \left( \frac{\partial u_i}{\partial x_j} + \frac{\partial u_j}{\partial x_i} - \frac{2}{3} \delta_{ij} \nabla \cdot \boldsymbol{u} \right) \tag{4}$$

for a kinematic viscosity $\nu$, where $x_i$ are Cartesian position coordinates and $\delta_{ij}$ is the Kronecker delta, i.e. the $3 \times 3$ identity matrix. We assume a constant $\nu$ within our simulation, set by the global $\alpha$-disc: $\nu = \alpha c_s H = \alpha c_s^2 / \Omega$. See Appendix A for a detailed discussion of this decision. Within the isothermal prescription $P = c_s^2 \rho$ where the sound speed $c_s$ is a constant. We have added to the Navier–Stokes equations a term $\boldsymbol{a}_{\text{SMBH}}$ to account for acceleration of gas due to the central SMBH inducing a rotating frame. This acceleration can be represented in terms of the angular velocity of the frame $\Omega$ and background shear parameter $q = \mathrm{d}\ln\Omega/\mathrm{d}\ln R$. As our frame is on a Keplerian orbit, we adopt $q = \frac{3}{2}$. With this prescription

$$\boldsymbol{a}_{\text{SMBH}} = 2\boldsymbol{u} \times \Omega \hat{\boldsymbol{z}} + 2q\Omega^2 \boldsymbol{x} \tag{5}$$

In this frame there exist equilibrium trajectories $\boldsymbol{u}_{\text{eq}}$ for which $\boldsymbol{a}_{\text{SMBH}} = 0$. These trajectories correspond to circular orbits about the SMBH in the non-rotating inertial frame.

$$\boldsymbol{u}_{\text{eq}} = \begin{pmatrix} 0 \\ -q\Omega x \end{pmatrix} \tag{6}$$

Furthermore, there is also contribution from the potential generated by the embedded black holes, which can be expressed as

$$-\nabla\phi_{\text{BH}}(\boldsymbol{r}) = \sum_{n=1}^{n_{\text{BH}}} m_{\text{BH},n} g \left( \frac{\boldsymbol{r} - \boldsymbol{r}_n}{h} \right), \tag{7}$$

where $\boldsymbol{r}_n$ and $m_{\text{BH},n}$ are the position and mass of the $n$-th BH, $n_{\text{BH}} = 2$, and $h$ is the softening length for the gravitational spline kernel $g(\boldsymbol{\delta})$

$$g(\boldsymbol{\delta}) = -\frac{G}{h^2} \boldsymbol{\delta} \begin{cases} \frac{32}{3}\delta - \frac{192}{5}\delta^3 + 32\delta^4 & 0 < \delta \leq \frac{1}{2} \\ -\frac{1}{15\delta^2} + \frac{64}{3}\delta - 48\delta^2 + \frac{192}{5}\delta^3 - \frac{32}{3}\delta^4 & \frac{1}{2} < \delta \leq 1 \\ \frac{1}{\delta^2} & \delta > 1 \end{cases} \tag{8}$$

Here $h$ is the gravitational smoothing length set to $h = 0.01 r_{\text{h, s}}$ as established in Section 3.

We do not include the effects of accretion in this study, in part due to the high uncertainties associated with the accretion methodology and rate. CR1 implemented a sink region which conditionally accretes SPH particles near the BHs, finding that accretion is a significant contributor to dissipation. In a follow up study, CR2 found that turning off this accretion prompts an increase in gas gravitational dissipation that compensates for the lack of accretion dissipation. Hydrodynamical studies often neglect a physical motivation for the accretion rate within sink regions, or else do not consider how inefficient angular momentum transfer may limit infall near the BHs. We expect that accretion rates close to the BHs should be closely tied to the viscous time-scale within these regions. As discussed in Appendix A, the true kinematic viscosity close to each BH is uncertain due to the delicate balance of gravitational and thermal effects. As such, we opt to remove any uncertainty introduced by poorly constrained accretion effects with intent to address accretion with a more developed model in future studies. If significant accretion rates are achieved in these systems, this may change the minidisc morphology and exert additional forces on the BBH (Li et al. 2023). Rowan et al. (2023b) found that binaries can form with or without the inclusion of accretion, though the source of torque differs in each case.

We adopt simple outflow boundary conditions in the $x$ direction (i.e. the radial direction with respect to the SMBH, see Fig. 1), but include a refilling routine at the $y$ boundaries. Refilling involves setting ghost cells to match the properties of the ambient disc. For outflow regions, all ghost cells are set to the values at the domain edge. No refill is imposed in the $x$ direction, as the viscous time-scale across the domain's radial extent is much longer than the simulation time. For the $y$ regions, we impose outflow in 'downstream' regions, and refill ambient gas in the 'upstream' regions.

$$(y = y_{\text{min}}) = \begin{cases} x < 0 & \text{refill} \\ x \geq 0 & \text{outflow} \end{cases} \tag{9}$$

$$(y = y_{\text{max}}) = \begin{cases} x \leq 0 & \text{outflow} \\ x > 0 & \text{refill} \end{cases} \tag{10}$$

These boundary conditions allow us to ensure minidisc formation does not artificially deplete the shearing box of gas.

## 2.2 Black hole dynamics

As each BH perturbs the gas around it gravitationally, they too experience feedback forces to consistently evolve them within the simulation. This is handled by a custom addition to the Athena++ code which tracks and updates the characteristics of $n$ BHs globally. To ensure that the BH trajectories are properly propagated, we adopt Quinn's method (Quinn et al. 2010), specifically designed to integrate equations of motion in the shearing frame where acceleration is a function of both position and velocity. During periods of close encounter, resolving the BH motion requires significantly smaller time-steps than indicated by the Courant–Friedrichs–Lewy (CFL) condition as the BH velocity exceeds that of the local gas. To better set this time-scale, we evolve the BHs and gas on a time-scale limited by the BH dynamics. This dynamic time-scale is set by the minimum of the combined flyby ($T_{\text{fb}}$) and free-fall ($T_{\text{ff}}$) time-scales for both the BH–BH pair, and the SMBH–BH pairs (e.g. Boekholt, Vaillant & Correia 2023b).

$$\Delta t = \eta \min \left( T_{ij} \right) \qquad T_{ij}^{-4} = T_{\text{fb},ij}^{-4} + T_{\text{ff},ij}^{-4} \tag{11}$$

$$T_{\text{fb},ij}^{-2} = v_{ij}^2 / r_{ij}^2 \qquad T_{\text{ff},ij}^{-4} = \left( G M_{ij} \right)^2 / r_{ij}^6, \tag{12}$$

where $M_{ij} = m_i + m_j$, $v_{ij} = |\boldsymbol{v}_i - \boldsymbol{v}_j|$, $r_{ij} = |\boldsymbol{r}_i - \boldsymbol{r}_j|$ are the total mass, relative velocity, and separation for BHs $i$ and $j$. $\eta$ is a scaling pre-factor set to 0.001 for stability; the selection method of this value is discussed in Appendix B. When the BHs are distant, the simulation evolves at a time-step set by the CFL condition and consideration of viscous time-scales as handled by Athena++ internally.

Our BHs are initialized within the simulation impulsively at $t = 0\,\text{yr}$ and do not experience gas gravitation until $t \approx 6\,\Omega^{-1} = 29\,\text{yr}$ when they have formed their own disc and streamers. This is done to reduce the dependence of capture on early disruption to the BH trajectory by asymmetries in the young unstable minidiscs. Each simulation will automatically terminate if a BH escapes the domain.

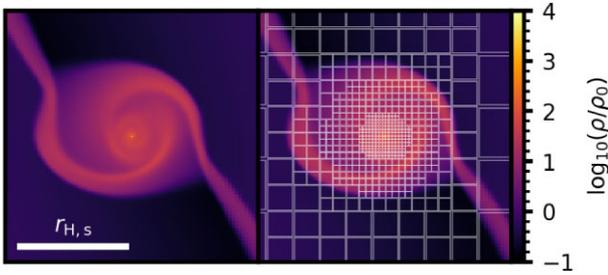

**Figure 2.** Logarithmic density plots of a BH minidisc before close encounter, with a side panel displaying the MeshBlock hierarchy. Each MeshBlock contains 16×16 cells: spatially smaller MeshBlocks close to the BH indicate an increase in resolution. There is a step down in refinement outside $0.2r_{H,s}$ and again at $0.5r_{H,s}$ where $r_{H,s}$ is the single Hill radius, see equation (1). Outside of $0.5r_H$, refinement decreases to the base level while preserving a maximum level difference of 1 between adjacent MeshBlocks.

BH properties (position, velocity, etc.) are logged every simulation time-step[1] providing temporally high-resolution data for analysis.

## 2.3 Adaptive mesh refinement

Constructing a meshed simulation that is large enough to give each isolated BH time to form a minidisc, while also fine enough to resolve the short length-scale dynamics at close encounter requires mesh refinement: changing the size of cells across the simulation. When using mesh refinement with Athena++, the simulation domain is decomposed into subsections called MeshBlocks, each containing the same number of cells. We implement an add-on to Athena++'s adaptive mesh refinement (AMR) routine, applying progressively higher levels of refinement the closer a MeshBlock is to a BH. Within $0.2r_{H,s}$ of a BH, the mesh is maximally refined at 6 levels above the root grid. Within the maximum refinement zone the cell size $\Delta x \sim 0.002r_{H,s}$. We apply one level less of refinement to MeshBlocks within $0.5r_{H,s}$ of a BH ($\Delta x \sim 0.004r_{H,s}$). This is demonstrated in Fig. 2 where each white outlined MeshBlock contains 256 cells. For a typical time-step before close encounter, this means that ∼ 65 per cent of all MeshBlocks are within $0.5r_{H,s}$ of a BH, equivalent to around $10^5$ cells. For a comparison to previous binary formation studies, Li et al. (2023) utilized a grid code with 50–100 cells per $r_H$ at maximum resolution; each of our simulations have over 500 cells per $r_H$ on the highest refinement level. This configuration allows for significantly lower computational expense whilst still adequately resolving the minidisc dynamics and chaotic flows at close encounter, with each individual simulation costing only 60–120 core hours.

## 3 INITIAL CONDITIONS

By restricting our study to a small region of the disc, we also simplify our initial conditions. To provide better comparison to CR1, we adopt similar global properties with a Shakura-Sunyaev disc of alpha viscosity $\alpha = 0.1$. The shearing box orbits the central SMBH of mass $M_{SMBH} = 4 \times 10^6 \, M_\odot$ at a radial distance $R = 0.0075$ pc, motivated physically by the pileup of binaries and frequent mergers at these radii observed in the semianalytical models of Tagawa et al. (2020a). In principle, we expect the outcome of gas assisted mergers to depend both on the mass ratios of the BHs involved, and the position of the interaction within the AGN disc.[2] Due to added computational expense associated with expanding the dimensionality of the parameter space, we limit ourselves to equal mass interactors with $m_{BH} = 25 M_\odot$ at a single radial distance from the SMBH. For this BH–SMBH mass ratio, we attain individual and binary Hill radii of $r_{H,s} = 10^{-4}$ pc and $r_H = 1.2 \times 10^{-4}$ pc. The simulation domain spans 0.0015 pc (12.5 $r_H$) radially and 0.003 pc (25 $r_H$) azimuthally: we then split this domain into a base grid of 128×256 cells (128 MeshBlocks). On top of this base mesh, we apply a further six levels of AMR, for seven levels total (see Section 2.3). In the shearing box, we set parameters that would vary in the global disc to constants, namely the disc scale height $H$, isothermal sound speed $c_s$, and kinematic viscosity $\nu$. For this study we set $H/R = 0.005$ to conform to a geometrically thin disc, such that for this stellar black hole mass $H \sim 0.3r_H$. This scale height then sets $c_s = H\Omega$, and $\nu = \alpha c_s H$. The embedded BHs are not massive enough to open gaps in the AGN disc (Kocsis, Yunes & Loeb 2011). The density of the ambient flow ($\rho_0$) is treated as a free parameter for this investigation, stated generally in units of $\rho_* \equiv M_{bin} r_H^{-3}$. While the AGN disc properties (aspect ratio, sound speed etc.) are generally consistent with geometrically thin models, the densities explored in this study are intentionally lower than predicted by models such as Goodman & Tan (2004). This decision was made to probe the extreme lower edge of successful binary formation. For the gas densities explored in this study, the disc is strongly stable against disc self-gravity ($Q > 1$) (Toomre 1964).

The background gas is initialized on a Keplerian orbit with respect to the SMBH, but its velocity field is warped near each BH to reduce chaos in the flows at early times when the BH is suddenly injected (see Appendix C). The BHs are inserted near the $y$ (or $\phi$) boundary of the shearing box on circular orbits. By starting our simulation when the BHs are distant, they are given enough time to form their own stable minidiscs before interacting. They are initially separated azimuthally by 22° and radially by an impact parameter $b$ which is varied between simulations. Throughout the study, $b$ is given in units of the binary Hill radius $r_H$. The BBH system is initialized such that its centre-of-mass lies in the centre of the shearing box, so as to maximize viable observation time for the interaction. We set the gravitational smoothing length for each BH to $h = 0.01r_{H,s}$. We note that gravitational softening is applied only to the BH–gas interactions, the BH–BH interactions feature no softening.

## 4 FIDUCIAL RESULTS

To better understand the dynamics of such complicated systems, we first examine a single fiducial model of a successful gas-assisted capture with initial radial separation $b = 2.3r_H$ and ambient density $\rho_0 = 1.9 \times 10^{-4} M_{bin} r_H^{-3}$. In this analysis we introduce a series of tools used to describe the hydrodynamic, orbital, and energetic evolution of the black holes and their disc environment. In Section 5 we expand the study of these systems by amending parameters in the fiducial model.

### 4.1 Evolution in isolation

Fig. 3 displays the structure of the BH discs before interaction. When the two are far apart, each BH forms its own accretion disc (minidisc). Due to Keplerian shear in the flow, these discs form spiral arms (streamers) that extend inwards and outwards radially: the inner arm leads the BH orbit, the outer arm lags behind. The BHs are massive enough to drive significant vacuation in the AGN disc away from these streamers, with $\rho \sim \rho_0/10$ in the cavities. Each BH initially travels on its equilibrium trajectory in sync with the ambient gas. To avoid perturbation from these trajectories while the discs are forming, we turn off gas gravitational effects on the BH until $t \sim 29$ yr. As they traverse the azimuthal simulation domain, the BHs gravitate towards each other and begin to distort each other's disc. The BH–BH interaction becomes more significant when they enter each other's mutual Hill sphere at $t \sim 36$ yr.

---

[1] Due to storage restrictions, the minimum time-step for BH logging is approximately $5 \times 10^{-7}$ yr.

[2] Per equation (5), BH interactions closer to the central SMBH will experience stronger frame forces, requiring more significant hardening to capture conclusively. Further study is required to determine the exact dependence of BBH formation on radial position in the AGN disc.

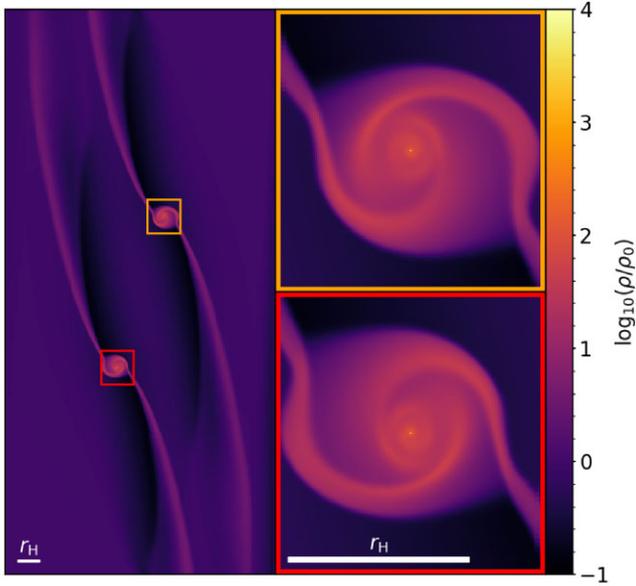

**Figure 3.** Logarithmic density plot of fiducial simulation at $t \sim 24$ yr with cutaways for each BH minidisc. Both BHs have formed stable gaseous minidiscs with defined spiral structure and strong overdense streamers that propagate across the entirety of the box. Outside the streamers, each BH drives a vacuation leaving underdense regions in the disc.

### 4.2 Evolution at capture

Fig. 4 displays the evolution of gas density in the vicinity of the two BHs as they approach each other. As the two bodies enter their mutual Hill sphere, their pull is strong enough to deform each other's minidisc and streamers. Once close enough, the streamers interior to the BHs coalesce, forming a bar of shocked gas connecting the two minidiscs. The BHs undergo a close encounter at $t \sim 40$ yr, passing through each other's discs and in doing so dissipating a substantial proportion of their energy to the surrounding gas, leading to the formation of a binary. By depositing orbital energy into the gas, the binary drives significant outflows that strip gas from the inner regions and eject it to the edges of the Hill sphere. There remains sufficient gas in/around the binary to drive further hardening during subsequent periapsis passages, though they are significantly weaker than the first hardening event. The binary components struggle to consistently maintain circumsingle minidiscs, as they are disrupted during each periapsis encounter and by interactions with the chaotic circumbinary flow. The evolution described above is consistent with that of Rowan et al. (2023b), figs 1 and 3).

### 4.3 Orbital dynamics

We explore the orbital properties of the binary by means of the eccentricity $e$ and centre-of-mass (COM) energy $E_{\rm bin}$.

$$e = \sqrt{1 + \frac{2E_{\rm bin}L_{\rm bin}^2}{G^2 M_{\rm bin}^2 \mu^3}}, \tag{13}$$

where we have introduced the binary mass $M_{\rm bin} = m_1 + m_2$, reduced mass $\mu = m_1 m_2/M_{\rm bin}$, and angular momentum $\boldsymbol{L}_{\rm bin} = \mu \, (\boldsymbol{r}_1 - \boldsymbol{r}_2) \times (\boldsymbol{v}_1 - \boldsymbol{v}_2)$. We define $E_{\rm bin}$ as the total energy of an isolated binary in its COM frame

$$E_{\rm bin} = \frac{1}{2}\mu |\boldsymbol{v}_1 - \boldsymbol{v}_2|^2 - \frac{GM_{\rm bin}\mu}{|\boldsymbol{r}_1 - \boldsymbol{r}_2|} = -\frac{GM_{\rm bin}\mu}{2a} \ . \tag{14}$$

The final equality applies if $E_{\rm bin} < 0$, where $a$ is the semimajor axis.

In isolation, for two bodies to form a bounded binary they must reduce their total energy such that $E_{\rm bin} < 0$. However when two bodies orbit a third one, the SMBH, the binary separation must remain bounded within $r_{\rm H}$ otherwise the tidal and Coriolis forces exerted on the binary may ionize it. The exact boundary for stability within hierarchical triples is dependent on the triple's properties, numerous studies have been made into the functional form of this boundary (Mardling & Aarseth 2001; Tory, Grishin & Mandel 2022; Vynatheya et al. 2022). Notably, retrograde binaries are expected to remain stable for larger semimajor axes than prograde (less negative energies), due to the direction of Coriolis acceleration. Within this study, we adopt a single stability threshold for all binaries, requiring that

$$E_{\rm bin} < \chi E_{\rm H}, \tag{15}$$

where $\chi = 2$ is a stability parameter set empirically based on the simulations of Tory et al. (2022) and

$$E_{\rm H} = -\frac{GM_{\rm bin}\mu}{2r_{\rm H}} = \frac{a}{r_{\rm H}}E_{\rm bin} \tag{16}$$

is the Hill energy: the total energy of a binary with semimajor axis of $a = r_{\rm H}$. This criterion is equivalent to requiring that the binary apoapsis $r_a$ remain within the binary Hill sphere for all eccentricities $e < 1$. We expect the relatively strict $\chi = 2$ criterion to correctly eliminate the vast majority of unbound systems, while only failing to recognize a small minority of bound systems.

Fig. 5 details the evolution of the fiducial binary orbital elements. We mark two times of interest at $t = 36$ yr and $t = 39.6$ yr. At the former, the binary energy becomes negative for the first time, indicating the formation of an unstable binary by SMBH tidal forces just after the two BHs enter their mutual Hill sphere. At the latter, a stable binary is formed during the first close encounter, with the binary energy rapidly decreasing due to gas gravitational effects. The binary energetic evolution is discussed in full in Section 4.4. The binary is highly eccentric, and remains so during the subsequent orbits. While the binary is hard, its high eccentricity results in a relatively large separation at apoapsis where the SMBH tides work most efficiently. This results in oscillations in angular momentum and eccentricity with the same period as the binary. In the following sections, we will discuss the mechanism for this binary formation and its subsequent evolution.

### 4.4 Dissipative effects

To distinguish what phenomena may be attributed specifically to gas dynamics as opposed to the complex behaviour expected of many-body systems, we must quantify how different forces interact

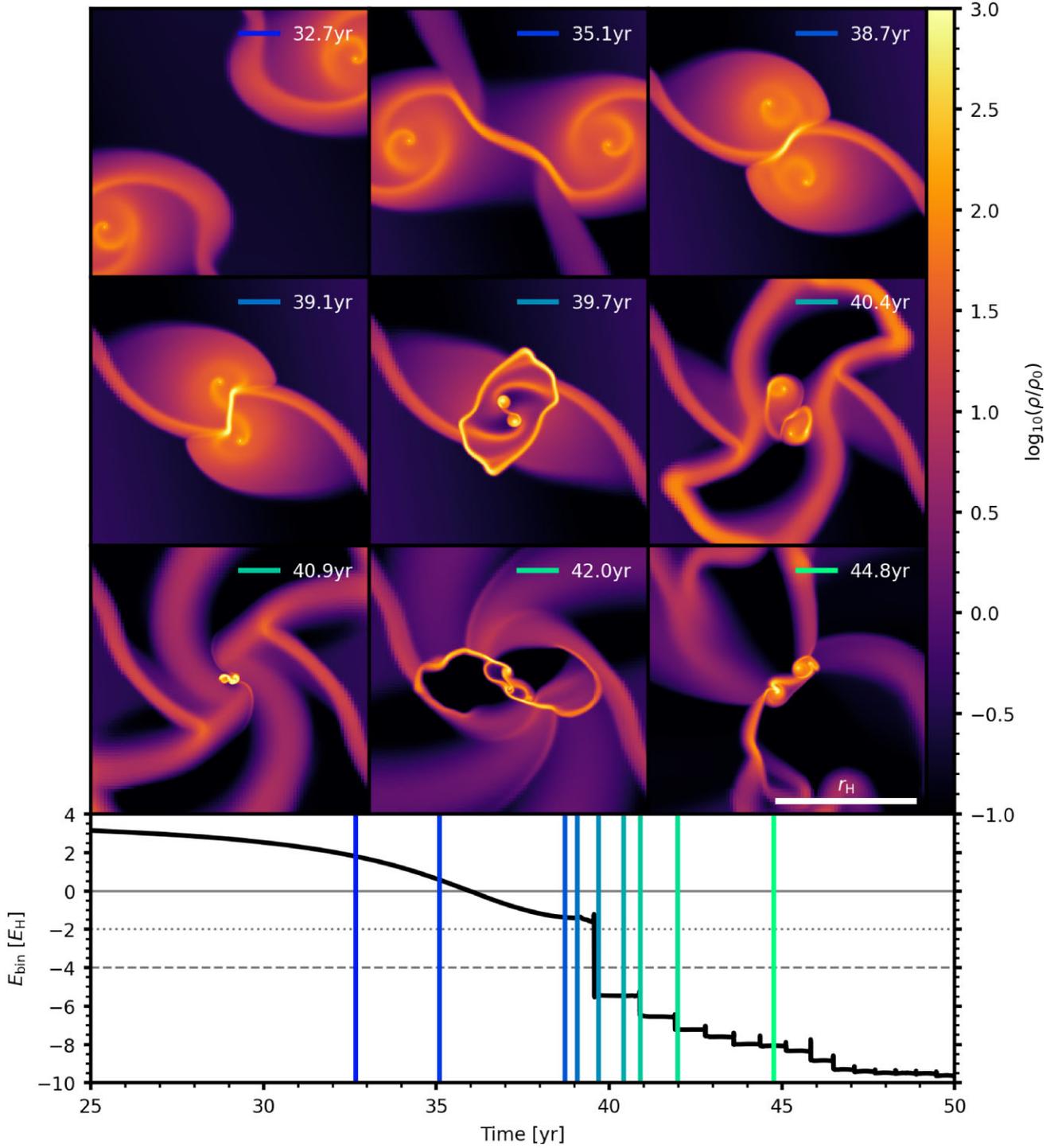

**Figure 4.** Grid of logarithmic density plots for the fiducial simulation at various snapshots from before first encounter to late in the hardening process. Lower panel shows the binary energy in units of the Hill energy $E_H$. As the BHs approach periapsis, the two minidiscs collide and form a shocked bar of gas. The binary forms during the first close encounter at $t \sim 40$ yr due to strong gas gravitational dissipation; orbital energy exchanged with the gas disrupts the BH minidiscs and partially depletes the mutual Hill sphere of gas. In subsequent periapsis passages, the binary undergoes further impulsive hardening events of lessening strength.

energetically with the binary. Work can be done on the binary due to:

(i) SMBH forces: $\epsilon_{SMBH}$ – each BH experiences centrifugal and Coriolis forces due to the rotating frame, we associate these forces with the central SMBH.

(ii) Gas gravitation: $\epsilon_{gas}$ – from differential gravitational gas attraction across the binary components[3]

---

[3]With accretion neglected, the only dissipation applied to the BBH as a result of gas effects is by BH-gas gravitation.

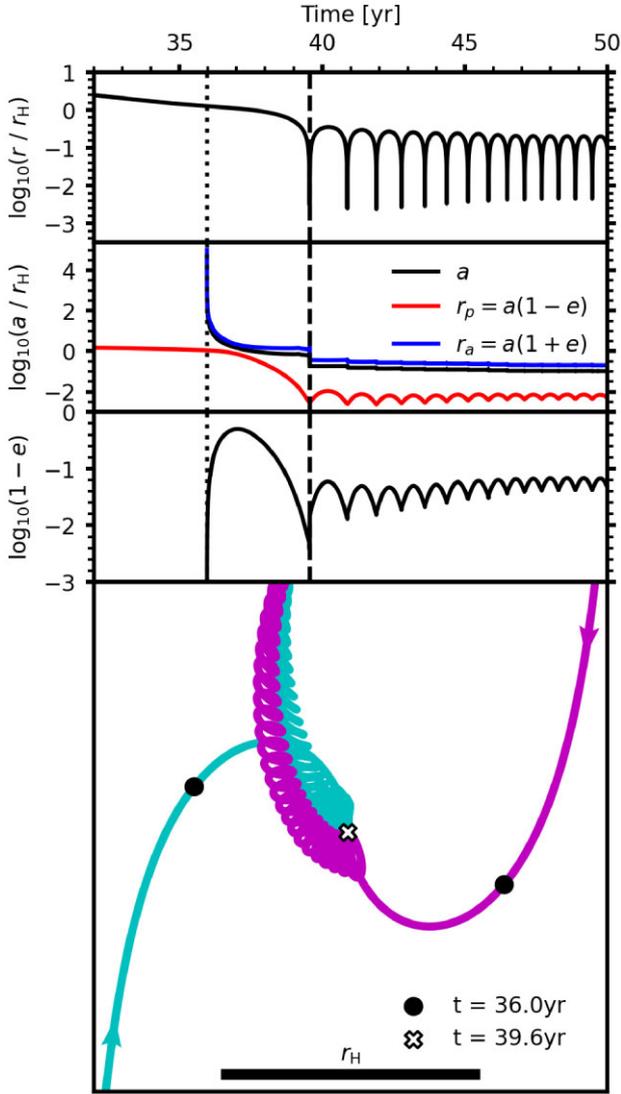

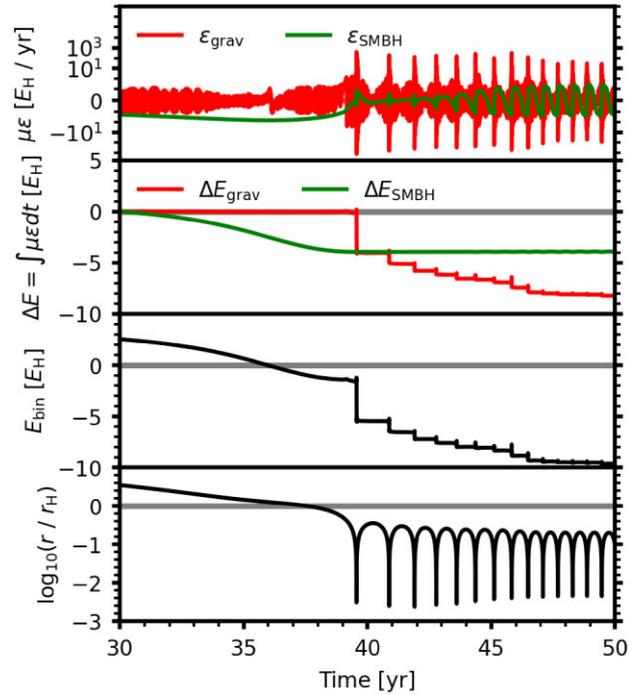

**Figure 5.** Evolution of orbital elements for the fiducial binary system, with the lower panel depicting the trajectories of each of the binary components in the shearing frame. Here $r$, $a$, $r_p$, and $r_a$ are the instantaneous binary separation, semimajor axis, periapsis, and apoapsis. The BHs undergo their first close encounter at 40 yr during which gas dissipation results in a stable binary forming. The young binary is knocked away from its equilibrium position and propagates through the shearing box, but the two BHs remain bound together. Torquing by the SMBH results in oscillations in the binary eccentricity on the same period as the binary. The vertical dotted and dashed lines at $t = 36, 39.6$ yr represent two points of interest: the first time the binary energy becomes negative, and the first periapsis, respectively. The positions of the black holes at these times are marked in the bottom panel.

Each of these measures can be calculated from considering the time derivative of the specific binary energy. The full derivation of this quantity is given by Appendix B of CR1.

$$\epsilon = \frac{\mathrm{d}}{\mathrm{d}t}\left(\frac{E_{\mathrm{bin}}}{\mu}\right) = (\boldsymbol{v}_1 - \boldsymbol{v}_2)\cdot(\boldsymbol{a}_1 - \boldsymbol{a}_2) \qquad (17)$$

We separate the acceleration corresponding to different physical processes as

$$\epsilon_{\mathrm{SMBH}} = (\boldsymbol{v}_1 - \boldsymbol{v}_2)\cdot(\boldsymbol{a}_{1,\mathrm{SMBH}} - \boldsymbol{a}_{2,\mathrm{SMBH}}), \qquad (18)$$

**Figure 6.** Energetic evolution of the fiducial model as the BHs approach first encounter. Upper panel displays variations in the respective binary energy dissipation rate due to the SMBH and gas gravity. The second panel integrates this dissipation over time (starting from $t = 30$ yr) to show the total change in binary energy, which is also plotted directly from the orbital elements in the third panel. The lower panel displays the separation of the binary components. It is clear that close encounters between the black holes are associated with impulsive dissipation events. The SMBH only contributes significant dissipation before the binary is formed at $t \sim 40$ yr, after which gas gravitation dominates.

where $\boldsymbol{a}_{\mathrm{SMBH}}$ is expressed in equation (5), and for the gas gravitation,

$$\epsilon_{\mathrm{gas}} = (\boldsymbol{v}_1 - \boldsymbol{v}_2)\cdot(\boldsymbol{a}_{1,\mathrm{gas}} - \boldsymbol{a}_{2,\mathrm{gas}}). \qquad (19)$$

Here $\boldsymbol{a}_{n,\mathrm{gas}}$ is calculated by summing the gravitational attraction on the BH $n \in \{1, 2\}$ by all $N_c$ gas cells in the simulation (with masses and positions $m_i$ and $\boldsymbol{r}_i$).

$$\boldsymbol{a}_{n,\mathrm{gas}} = G\sum_{i=1}^{N_c} m_i \frac{\boldsymbol{r}_i - \boldsymbol{r}_n}{|\boldsymbol{r}_i - \boldsymbol{r}_n|^3} \qquad (20)$$

Fig. 6 details the energetic evolution of the fiducial model with these measures. On approach, energy is steadily extracted from the binary by the SMBH, with gas gravitation being subdominant. Gas gravitation dissipation peaks during the first close encounter, reducing $E_{\mathrm{bin}}$ by $\sim 4E_{\mathrm{H}}$ and forming a stable binary. The SMBH continues to do work on the binary, but on smaller magnitudes than gas gravitation dissipation and secularly the SMBH's energetic effect averages out to approximately zero once the binary is formed. Gas dissipation continues to peak during subsequent close encounters. The energetic evolution of the bound system follows a clear pattern: impulsive dissipation events at each periapsis harden the binary. As such, the binary energy evolves as an approximate step function, with hardening periods at periapsis but only minor energetic evolution outside of these events. The frequency of these events increase as the binary period shortens, creating a system that rapidly hardens due to gas effects. The following periapsis passages are similarly deep and high velocity to the first, but dissipate significantly less energy: our simulations show that this is likely due to a reduction in gas mass local to the binary as generated by strong outflows during the first periapis.

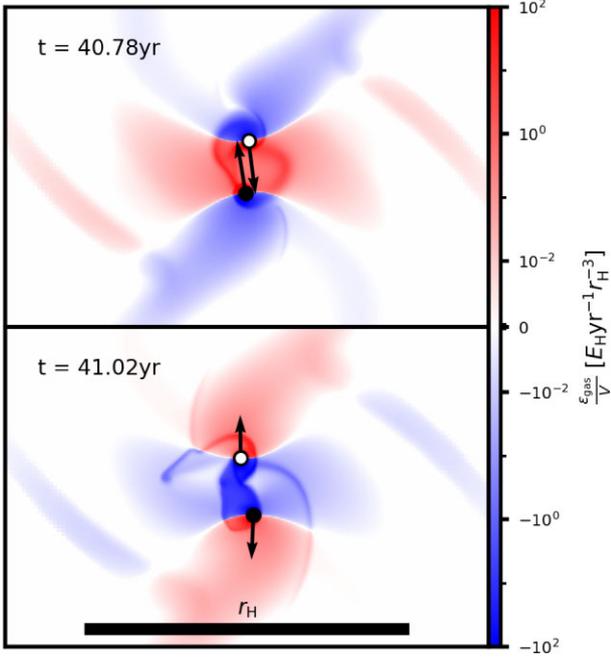

**Figure 7.** Spatial distribution of gravitational gas dissipation density, for the fiducial model before and after a periapis passage. Highlighted in red are regions of gas injecting energy into the binary (gas leading the orbit), in blue energy is extracted (by gas lagging the orbit). Energy is generally injected on in-fall due to the extra gas mass in the binary system, but extracted as the binary travels towards apoapsis due to minidisc gas being unable to keep up with the binary components. The net effect is a removal of orbital energy from the binary.

The strong localization of dissipation to periapis passages can be explained by consideration of the BBH dynamics, the dissipation equation, and the spatial dissipation density distributions. First, for equivalent gas acceleration fields, we expect BBHs with greater relative velocities to dissipate more energy [in equation (17), velocity is a linear term]. We might then question why there is a differential acceleration across the binary components near periapis. This can be better understood by considering the distribution of gas near the BHs during close encounters. When in isolation, the minidiscs are well bound to the BHs and travel with them through the AGN disc. The minidiscs collide during close encounters, with the BHs accelerating ahead of their bound gas as they do not feel the same hydrodynamic forces (pressure, viscosity) which generally retard the gas on collision. This leads to a build up of gas behind the BHs and results in a drag force that removes energy from the binary. Fig. 7 displays the spatial distribution of dissipation density, where blue regions indicate gas extracting energy from the binary, and red injecting into it. While on in-fall energy injection often exceeds extraction, mass stripped from the minidisc during collision leads to a strong drag after periapis that results in a net loss of energy. These dissipation events are very short for deep, eccentric flybys as the binary components spend very little time close to periapis where the gas gravitation is strongest. As such, we observe different dissipation chronologies to GDF studies such as DeLaurentiis et al. (2023) which found capture outcome to be determined before close interaction, whereas here the outcome is determined by gas effects very close to periapis. This difference in chronology is likely due to GDF decreasing with increased velocity for supersonic objects, resulting in lower drag for deeper, faster periapis passages. Our chronology and spatial distribution of dissipation are consistent with Li et al. (2023): during each period of the binary, dissipation is strongest immediately after each periapis. Cumulatively, more energy may be dissipated in the subsequent periapis passages than in the first, though this varies between runs. Neglected from this simulation are the effects of gas self-gravity; it is unclear whether introducing an extra acceleration term to the gas would limit or enhance the lagging which drives gas dissipation.

### 4.5 Torques and angular momentum

We can attribute specific torques to the physical processes in a similar method to the definition of dissipation rates in Section 4.4.

$$\tau = \frac{d}{dt}\left(\frac{L_{bin}}{\mu}\right) = (\boldsymbol{r}_1 - \boldsymbol{r}_2) \times (\boldsymbol{a}_1 - \boldsymbol{a}_2) \tag{21}$$

$$\tau_{SMBH} = (\boldsymbol{r}_1 - \boldsymbol{r}_2) \times (\boldsymbol{a}_{1,SMBH} - \boldsymbol{a}_{2,SMBH}) \tag{22}$$

$$\tau_{gas} = (\boldsymbol{r}_1 - \boldsymbol{r}_2) \times (\boldsymbol{a}_{1,gas} - \boldsymbol{a}_{2,gas}) \tag{23}$$

Analogous to the Hill energy $E_H$, we introduce $L_H$ as the angular momentum of a circular binary with $a = r_H$, such that $L_H = \mu\sqrt{GM_{bin}r_H}$.

Fig. 8 details the angular momentum evolution of the fiducial system under these torques. When the BHs are far apart the angular momentum is large and negative with respect to the AGN disc rotation. As the simulation progresses, the angular momentum becomes less negative, driven primarily by the SMBH. There is a $e < 1$ epoch before true capture during which the BHs are in an unstable binary ($E_{bin}/E_H < 2$). Following the first encounter when the binary becomes tightly bound, the angular momentum and energy evolution are remarkably different. While the energy continues to decrease during subsequent passages, the angular momentum is nearly constant. Torques by gas gravitation are largely insignificant throughout the evolution, but we might expect gas torques to dominate at late times when the binary has accumulated further gas and SMBH have diminished due to hardening. The binary remains highly eccentric after the first encounter, the eccentricity oscillates between 0.9 and 0.99. Over suitably long time periods, dissipation at periapsis may damp eccentricity and circularize the binary (Li et al. 2023), but this is not observed during the time-frames of this study. As the semimajor axis and apoapsis shrink, the periapsis distance is nearly constant $r_p \sim 0.003 r_H$ during the evolution. This type of behaviour is opposite to the hierarchical triple evolution without gas, e.g. the eccentric Lidov–Kozai mechanism (Naoz 2016), where $r_p$ changes at nearly constant $a$ and it is reminiscent of the evolution for dissipative short-range-interaction processes in highly eccentric binaries such as tidal capture or GW capture.

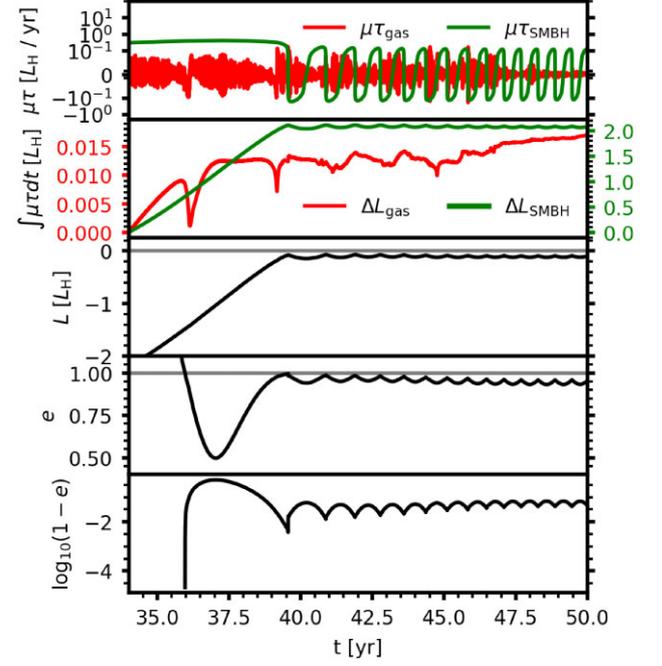

**Figure 8.** Angular momentum evolution for the fiducial model starting from when the black holes are $1.5r_H$ apart, with panels for instantaneous torques, integrated torques, total binary angular momentum, and eccentricity. The angular momentum evolution is dominated by the SMBH throughout.

### 4.6 Summary of fiducial model

The fiducial model builds on previous binary capture studies to provide an independent confirmation for gas-assisted binary formation with different methodologies and systematics. Critically, we extended our model to greater refinement than previous grid-based studies, allowing us to better resolve the chaotic flow structure present in the gas during BH–BH close encounter. The capture is achieved by strong gravitational gas dissipation during the first close encounter. This binding injects substantial energy into the surrounding gas, generating chaotic outflows that strip gas from the minidiscs. The binary formed is highly elliptical, and remains so during its subsequent orbits. The SMBH dominates the torque on the binary, forcing oscillations in eccentricity. Post formation binary periapsis passages are accompanied by further hardening events, albeit with minor energy decreases compared to the initial close encounter.

## 5 PARAMETER STUDY

Here we will consider a grid of 345 models generated by amending the initial conditions of our fiducial model (see Section 4). For our base grid, we span the initial ambient disc density approximately uniformly in log space within $\rho_0 \in [3.3 \times 10^{-5}, 3.3 \times 10^{-3}] M_{bin} r_H^{-3}$ and impact parameter linearly within $b \in [1.3, 2.5] r_H$. We note a difference here to the range of parameters used in CR1: in that paper $r_H$ was the single BH Hill radius, as opposed to the binary Hill radius in this work (equation 1). This paper also considers ambient disc densities considerably lower than those analysed in CR1 which were derived from the Goodman & Tan (2004) AGN disc model. This is done to better probe the low-density limit of successful binary capture. We note that due to the linearity in density of the isothermal fluid equation (2), discs of varying initial densities are morphologically similar. As such the principle difference between models of different densities is the strength of the gravitational attraction on the BHs. This would not be the case for more general equations of state, where this linearity in density is not preserved. Compared to previous hydrodynamical studies that generally utilize a few to tens of models (Li et al. 2023), our wide and finely sampled parameter space allows us to better chart the complicated and non-linear nature of the BBH capture environment. The principal 'outcome' for each of these simulations concerns the success or failure of binary formation. Achieving a negative binary energy is insufficient to determine long-term binding, as SMBH tidal forces may yet ionize the system. A binary capture is considered successful if the binary energy $E_{bin}$ is sufficiently negative. As addressed in Section 4.3, the exact boundary for stability is complex, and in this work we take any binary achieving $E_{bin} < 2E_h$ to be stable (see equation 15). Each simulation terminates once it is sufficiently bound, has become unbound after its initial encounter, or if it reaches $t = 58$ yr: this is done to reduce computational cost.[4]

As such, these models provide lots of information about the nature of binary formation, but are not intended to be used for extended study of the resulting binary systems over many orbital periods. This is in part due to the cost of resolving hard binaries, where increasingly small time-steps are required to resolve the BH trajectories accurately (see Appendix B). We expect there to be a minority of cases where a binary identified as unbound at simulation stop may yet harden sufficiently due to gas dissipation to remain bound. Similarly, the empirically derived boundary of $\chi = 2$ is set by consideration of a large proportion of systems remaining bound beneath this energy; a very small minority of cases may only be quasi-stable. Despite this, we are confident that this observation period allows us to correctly identify the vast majority of binary systems.

### 5.1 Gas-free interactions

Boekholt et al. (2023a) explored the complex orbital dynamics of a flyby interaction in proximity to a massive companion and identified impact parameters for which arbitrarily close encounters between the flyby bodies could occur. Our initial conditions are slightly different, integrating from an azimuthal separation of $22°$ instead of a full $180°$ due to the limited extent of the shearing box. Despite this, we identify many of the same trends, specifically the existence of direct impact trajectories and orbit families associated with different impact parameters. We include here a brief study of the gas-less flyby scenario as test of our integrator and to help inform our understanding of the more complicated response of the system when gas is included. The relationship between impact parameter and periapsis of the first encounter in the gas-less case follows a double-valley shape, with the closest first encounters occurring at $b \sim 2.0 r_H$ and $2.3 r_H$, respectively. At these impact parameters, the trajectories of the BHs before close encounter lead to maximal angular momentum extraction by the SMBH, resulting in the deepest encounters. For sufficient fine tuning in $b$, one can find a trajectory where the BHs pass arbitrarily close to each other. For the initial conditions used in this study, we find these 'zero angular momentum' (ZAM) seperatrices near $b_1 = 1.9975 r_H$ and $b_2 = 2.3225 r_H$. When crossing a seperatrix, the sign of angular momentum at periapsis inverts. As such these trajectories mark the edges of families in encounter types, which we label prograde interior, retrograde interior, and prograde exterior (PI, RI, and PE, respectively). Fig. 9 describes the nature of these families in the gas-free case, within a frame centred on the inner BH. For $b < b_1$ (type PI), the outer BH will pass interior to the inner BH with positive angular momentum (prograde with respect to the AGN disc). If the impact parameter is increased slightly, such that $b_1 < b < b_2$ (type RI), the BH will still pass interior to the inner BH, but now with negative angular momentum (retrograde). Increasing to $b > b_2$ (type PE), the outer BH will remain exterior to the inner BH at periapsis, on a prograde orbit. For sufficiently small or large impact parameters, the BHs will fail to penetrate their mutual Hill sphere, and no close encounter will occur: for our initial conditions, we only observe mutual Hill sphere penetration for $1.6625 r_H \leq b \leq 2.4375 r_H$ in gas-less systems. When gas is introduced to the system, the exact position of these boundaries will move, but understanding the rough relationship between impact parameter, angular momentum extraction by the SMBH and therefore periapsis depth is crucial to interpreting the dissipative effects in the gaseous case, as the depth of first close encounter proves to be strongly correlated with binary formation.

---

[4]For systems that reach sufficient binding energies or become unbound after first encounter, the system is evolved for a further $\Delta t = \Omega^{-1} \sim 4.84$ yr to ensure that this is the end state of the system.

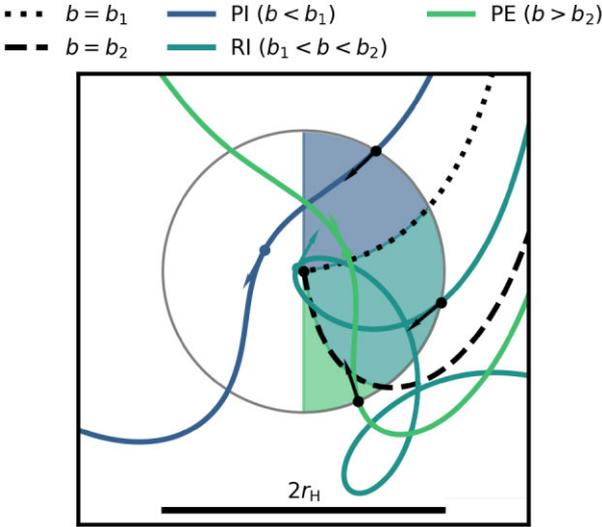

**Figure 9.** BH trajectories in the gas-less case: from a frame centred on the inner BH, each path follows the outer BH. Classes of encounters are identified (prograde internal, retrograde internal, and prograde external) by the initially external BH passing inside or outside the inner BH at periapsis, and in which direction w.r.t the global AGN disc rotation. Each encounter class is separated by a 'direct impact' trajectory, at $b_1 = 1.9975 r_H$ or $b_2 = 2.3225 r_H$. For trajectories with $b \lesssim 1.6625 r_H$ and $b \gtrsim 2.4375 r_H$ we do not expect close encounters to occur in gas-less systems.

Note that even in the absence of gas, impact parameters close to $b_1$ and $b_2$ lead to extremely close encounters that may result in binary formation due to GW emission (or tidal dissipation for stellar encounters): they may even lead to direct head-on collisions in the most fine-tuned case. Boekholt et al. (2023a) describes Jacobi encounters within hierarchical triples as systems where the inner binary experiences multiple close encounters driven by outer body, finding that these encounters occur for specific values of initial impact parameter defined as fractal regions A, B, C. Similar binary capture and direct collision trajectories also exist for subsequent encounters following the first encounter (within regions A, B, C), which increase the total binary capture cross-section cumulatively by approximately 25 per cent (Boekholt et al. 2023a). As our initial conditions differ from Boekholt et al. (2023a) the exact boundaries of the Jacobi regions are hard to identify, but we can mark their rough positions with respect to the ZAM trajectories. The first wide region A is found at $b$ values slightly beneath the lower ZAM trajectory at $b \sim 2.0 r_H$. The next region B is the widest and overlaps with the ZAM trajectory at $b \sim 2.3 r_H$. The final region C is considerably narrower than the other two and can be found at $b$ values slightly larger than the $b \sim 2.3 r_H$ ZAM trajectory.

### 5.2 Outcomes

Fig. 10 displays the full grid of 345 simulations with gas, spanning a range of initial ambient disc densities and BH impact parameters. A successful binary formation is shaded green, with a failed capture in orange. We note first the high number of captures in this parameter space, despite the relatively low disc densities represented. We see a high number of captures for initial ambient disc densities in excess of $4 \times 10^{-4} M_{bin} r_H^{-3}$. However, within a narrow band of impact parameters around $b = 2.3 r_H$, we are able to observe successful BBH formation at disc densities of only $\rho_0 = 10^{-4} M_{bin} r_H^{-3}$. We might expect a second narrow band to be present around $b = 2.0 r_H$ due to the deep close encounters predicted here, this discrepancy is discussed in Section 5.4. Secondly, we note that knowledge of either $\rho_0$ or $b$ is insufficient to determine the outcome of a close encounter on its own. Thirdly, we observe some failed captures appearing at the highest gas densities and smallest impact parameters (bottom right corner in the figure), as well as a very wide capture section at intermediate densities. These features are discussed in Section 5.5. Understanding the shape of this outcome space requires consideration of the dependence on energy dissipation on both the macro-disc properties and the BH trajectories. While we observe that the binaries formed are generally highly eccentric ($e \gtrsim 0.75$), we do not attempt to draw conclusions as to the eccentricity (or semimajor axis) distributions of the resulting BBHs. The captured BBHs continue to dynamically evolve, with semimajor axes that shrink periodically due to continual gas hardening at periapsis and eccentricities that oscillate with the binary period due to the SMBH (see Section 4.5).

### 5.3 Energetics at first close encounter

Fig. 11 plots the energy changes during the first encounter between BHs in each of the simulations in the grid. We calculate the energy change during the first encounter by integrating the dissipation from when the BHs first enter their mutual Hill sphere ($r = r_H$). When considering the energetics at periapsis, we separate effects attributable to the gas and to the SMBH. For gas dissipation, we integrate through to either the first apoapsis, or till the BHs escape the Hill sphere. For SMBH dissipation, we only integrate down to first periapsis. These ranges allow for the best predictions of energy soon after first close encounter; gas effects are more localized to the periapsis, whereas the rate of work done by the SMBH weakens as the BH separation decreases within the Hill sphere. We distinguish the data by the depth of first periapsis $r_p$ and ambient disc density $\rho_0$ (by marker size), separating the type of dissipative effect in the 2nd and 3rd panels. SMBH dissipation is a function of BH trajectory and therefore initial impact parameter, but at high disc densities the gas gravity may be strong enough to perturb the BHs onto different trajectories and so will also have an effect. When considering the relationship between energy dissipated and binary formation outcome, we note that only simulations that dissipate a significant amount of energy during their first close encounter result in successful binary formation, the threshold for this is approximately $O(E_H)$. Binaries that do not dissipate enough energy during the first flyby tend to be quickly ionized by the SMBH.

We see that generally, gas gravitation dominates the dissipation but SMBH effects can be significant at lower densities. At first glance, there appears to be little correlation between periapsis depth and energy dissipated. Finding a clear relationship between SMBH dissipation and periapsis depth is difficult, as SMBH dissipation is strongly related to the BH trajectory which may change significantly for small perturbations to the initial impact parameter. Furthermore, the relationship between the gas dissipation and periapsis depth is convoluted by the effect of changing disc density. Focusing only on dissipation by gas, we fit a power law to the energy dissipated during

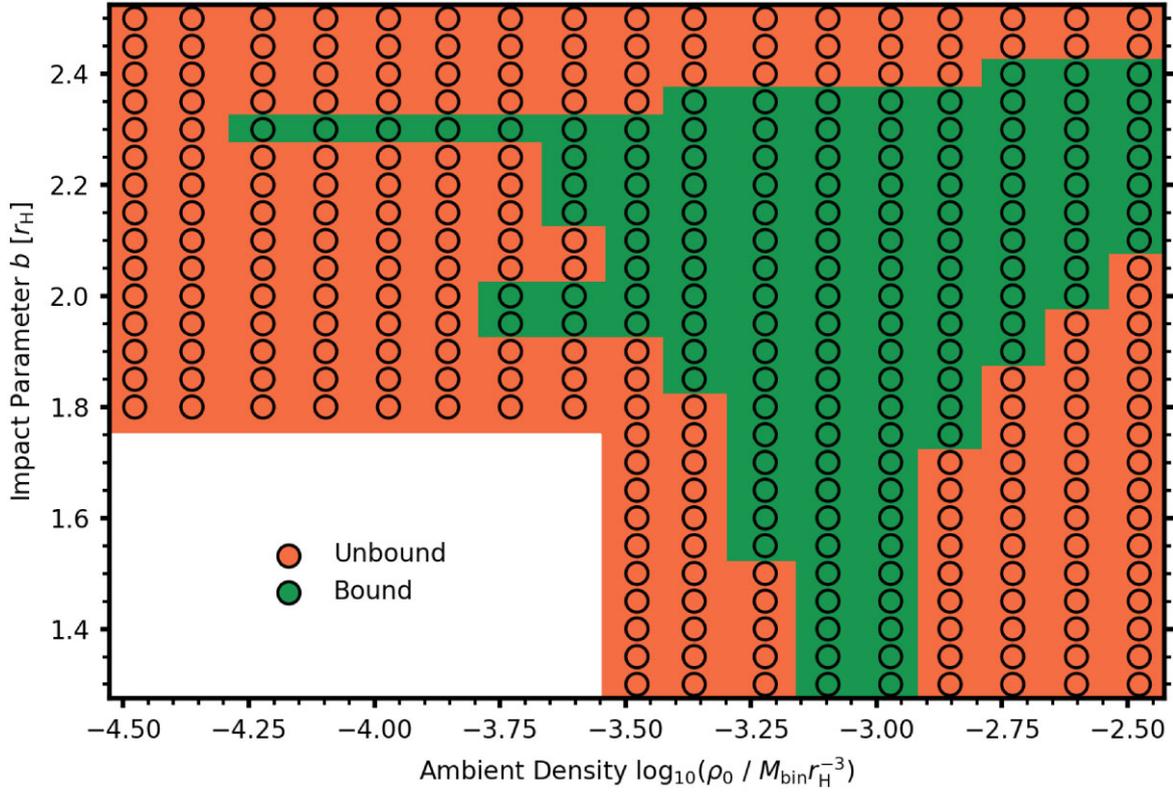

**Figure 10.** Encounter outcomes for BBH with varying initial ambient gas density and initial impact parameter. The 345 simulations are represented by circles, with outcomes colour coded. In the lower density limit, captures are restricted to those impact parameters which result in very deep first encounters with SMBH assist. As density is increased, captures start to occur on a wider range of impact parameters. At very high densities, pre-encounter gas effects induce a drift which reduces the capture cross-section (see Section 5.5).

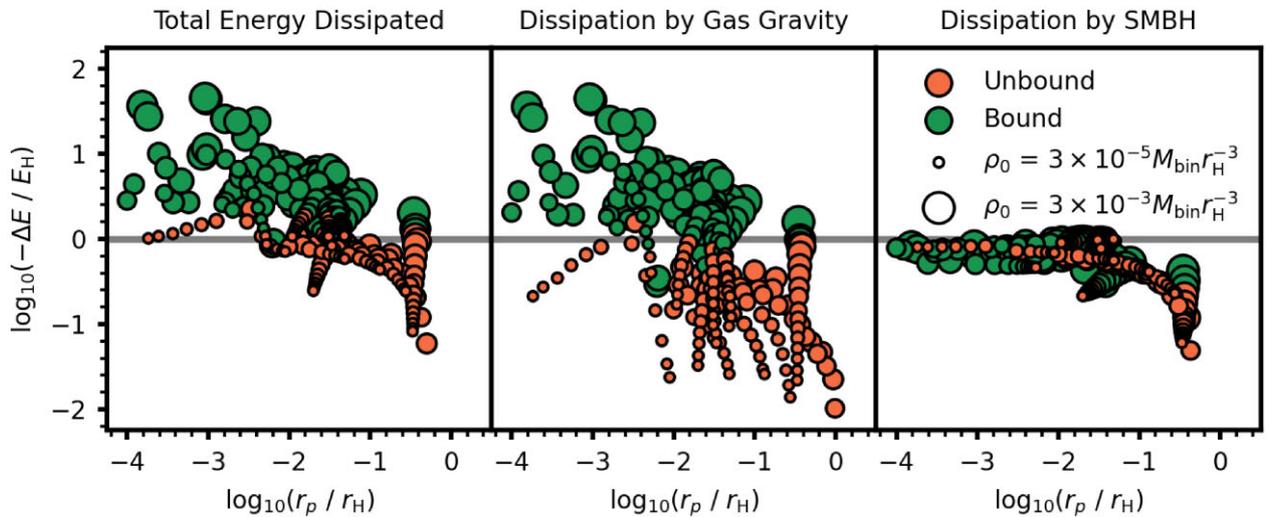

**Figure 11.** Absolute energy changes during the first close encounter separating total energy dissipation, gas gravitational dissipation, and SMBH dissipation from left to right. Each simulation is colour coded by its outcome with successful bindings in green and failed in orange. The ambient AGN disc densities $\rho_0$ are distinguished by marker size, with denser discs given larger markers. SMBH dissipation is bound beneath $\Delta E \sim E_H$, whereas gas dissipation ranges from $10^{-2}$ to $10^2 E_H$. Only runs that dissipate enough energy during the first close encounter (around $1-2E_H$) result in binary formation.

first periapsis, such that

$$\left| \frac{\Delta E_{gas}}{E_H} \right| = A \left( \frac{\rho_0}{M_{bin} r_H^{-3}} \right)^\alpha \left( \frac{r_p}{r_H} \right)^\beta \tag{24}$$

Table 1 lists the best-fitting parameters, with Fig. 12 displaying the results of this fitting, along with the residuals. We exclude the results of high gas density runs with $\log_{10}\left(\frac{\rho_0}{\rho_*}\right) > -2.8$ runs from the fit due to their significant deviations from the model, suggesting a decoupling at higher densities. The relationship is effectively linear in $\rho_0$, i.e. $\alpha = 1.01 \pm 0.04$, which we might expect; the mean density and mass of the minidiscs prior to the encounter are approximately proportional to $\rho_0$ and the gas–BH interaction is approximately linear in these quantities, while non-linear effects such as polarization and turbulence are apparently subdominant. The simplifying assumptions in the simulation (equations 2–4) may lead to this outcome: pressure is simply proportional to density, the sound speed, and kinematic viscosity are globally fixed and are not allowed to vary in time, and the gas self-gravity is neglected in this study. As discussed in Section 4.4, we expect dissipation to increase for faster, deeper flybys due to the greater disruption of the gas minidiscs, though the exact origin of the $\beta = -0.43$ exponent is harder to identify. Were $\Delta E$ entirely linear in relative BH velocity, we might expect $\beta \sim -0.5$ to match the free-fall solution with eccentricity $e$, $v_p = [GM_{\rm bin}(1+e)]^{1/2}r_p^{-1/2}$. On the other hand, faster flybys will naturally feature shorter periods where dissipation is significant, i.e. $t_p = r_p/v_p \propto r_p^{3/2}$ for nearly parabolic encounters, so it is not immediately clear that faster, shorter flybys should dissipate more energy overall. However, faster flybys are observed to induce more BH-minidisc lag which we expect to be the driving force for differential acceleration across the BH components (as presented in Fig. 7). Furthermore, when comparing relatively shallow encounters, smaller $r_p$ values will result in the BHs penetrating deeper into the higher density central sections of each other's minidiscs, leading to greater gravitational acceleration by gas. Thus we are left with a balance of competing factors that result in a net empirical dependence of $\Delta E \sim r_p^{-0.43}$.

**Table 1.** Best-fitting parameter values and dispersion ($\sigma$) for the orbital energy dissipated by gas in equation (24).

| Variable | Value | $\sigma$ |
|---|---|---|
| $\log_{10}(A)$ | 2.63 | 0.17 |
| $\alpha$ | 1.01 | 0.04 |
| $\beta$ | $-0.43$ | 0.03 |

There remains a high degree of scatter in the results about the modelled values, and it is clear that periapsis depth and disc density are not truly independent values (see Fig. 13). Moreover, the model regularly overestimates dissipation in the high density limit where $\log_{10}\left(\frac{\rho_0}{\rho_*}\right) > -2.8$, suggesting perhaps a levelling off in the scaling with density. Regardless, the fit provides a reliable order-of-magnitude estimate for the energy dissipated by gas during the first encounter, a fact we will use in Section 6 for our modelling.

We note here that while we have used $\rho_0$ to fit our model, it acts only as a proxy for minidisc mass which ultimately dictates the intensity of gas dissipation. If the system were left to evolve for longer before close encounter, the minidiscs would attract more gas and grow in mass. We approximate the minidisc mass $M_d$ as the mass of gas within $r_H$ of a BH. The linearity of density in the isothermal fluid equations (2) means that the minidisc mass is linearly dependent on the initial ambient density. As all our models evolve for a similar time before close encounter and feature BHs of equal mass we can predict $M_d$ just before close encounter from the ambient density (specifically, when the two BHs are $2r_H$ apart).

$$\left.\frac{M_d}{M_{\rm bin}}\right|_{r=2r_H} \sim 6\left(\frac{\rho_0}{M_{\rm bin}r_H^{-3}}\right) \qquad (25)$$

This relationship holds strong for the majority of models in the grid, for the details see Appendix E. However, this relationship is specific to our simulations, it is dictated by our assumptions. Indeed, it is likely that the Hill spheres are not saturated with gas mass and evolving the system for longer before the dynamical encounter would allow for more mass accumulation in the minidiscs.

Nevertheless, using the actual minidisc masses rather than the initial ambient density, the amount of energy dissipation by gas may be estimated more generally. We note that for non-isothermal systems, more massive minidiscs will likely be morphologically dissimilar to their lower density counterparts due to the minidiscs heating above the ambient disc. This would break the linear growth of minidisc mass with time and may have a significant effect on the disc radial structure, but recovering such results would require full simulation with a non-isothermal equation of state, beyond the scope of this work.

Curiously, we note that the measured total BH energy dissipation during the first encounter due to gas is consistent with the expression

$$\Delta E_{\rm gas} = 4.3 M_d v_H v_p, \qquad (26)$$

where $v_H = (GM_{\rm bin}/r_H)^{1/2}$ and $v_p = (2GM_{\rm bin}/r_p)^{1/2}$ are the BHs' relative velocity for a parabolic orbit at separations of $2r_H$ and $r_p$, respectively, and $M_d$ is the mass of a single BH minidisc at BH separations of $2r_H$ (i.e. equation 25 in our simulations). Indeed, equation (26) may be written equivalently as

$$\left.\left|\frac{\Delta E_{\rm gas}}{E_H}\right|\right|_{\rm toy\ model} = 293\frac{\rho_0}{M_{\rm bin}r_H^{-3}}\left(\frac{r_p}{r_H}\right)^{-0.5}, \qquad (27)$$

which is reminiscent of equation (24) and Table 1. If the fit to the simulations is done by forcing $\beta = -0.5$ in equation (24), then $A = 295^{+121}_{-86}$ (equations 26–27). Clearly, the dissipation rate calculated from GDF, $\Delta E_{\rm GDF} \sim G^2 M_{\rm bin}^2 \rho_0 v_{\rm BH}^{-2} r_H = G^3 M_{\rm bin}^3 \rho_0 v_{\rm BH}^{-2} v_H^{-2}$ (see also Tagawa et al. 2020a; DeLaurentiis et al. 2023; Rozner et al. 2023), is inconsistent with the scalings shown by our simulations (cf. equation 26). Equation (26) may be useful to build a physical model to explain the measured BH energy dissipation in the simulations. We develop a more accurate predictive model on the outcome of the dynamical encounter in Section 6.

### 5.4 Asymmetry at low gas density

Knowledge that the binary formation outcome is coupled to periapsis depth helps define the shape of the outcome shape presented in Fig. 10, specifically the low density protrusions at $b \sim 2.0r_H$ and $2.3r_H$. As explored in Section 5.1, these impact parameters result in the deepest first encounters in the gas-less limit. This is also the case at low gas densities, as depicted in Fig. 13. However, we might question why the tail at $b = 2.3r_H$ extends to lower densities than the $b = 2.0r_H$ branch. There is a fundamental difference between the dynamics in the two valleys: not only does the upper valley ($b \sim 2.3r_H$) have a deep first encounter, but it also has multiple encounters even in the gas-less case. This is because the $2.3r_H$ valley overlaps with Region B of Boekholt et al. (2023a) where Jacobi captures occur. Conversely, the valley at $b = 2.0r_H$ does not lie within a Jacobi region: despite the deep first encounter there is no guarantee of a second encounter at low density. While we generally find the first close encounter to be the most significant, we might conclude that the reason why the $b = 2.0r_H$ valley does not extend to lower densities is because they lack the subsequent close encounters observed in the $b = 2.3r_H$ valley. The BH trajectories in the $b = 2.0r_H$ case feature very deep close encounters, but after this the BHs rapidly separate. This might indicate a further dependence within the gas dissipation equation (24), where we should also consider the precise orbital trajectory as well as depth and density. Alternatively, this may be a sampling artefact, with a narrow valley extending deeper at $b = 2.0r_H$ but beneath our sampling resolution.

We do expect gas capture to be possible within all of the Jacobi encounter regions, as there exist arbitrarily deep encounters in the subsequent close encounters within those regions due to the underlying fractal structure. However, the width of these subregions are incredibly narrow and beneath our sampling resolution in $b$. The only region obvious in our system is Region B, as it also lies within the valley of deep first close encounter and as such binary formation is easier there. We do observe some members of Region A at small $b$, which do not form binaries but do feature multiple encounters (see Section 5.6). There exists also a very narrow Region C for Jacobi encounters at large $b$, and while we do observe some looping behaviour around $b \sim 2.45$, we do not observe multiple close encounters (where $r_p < r_{\rm H}$). There presumably exist Jacobi encounters near this $b$ value which are not sampled by our grid.

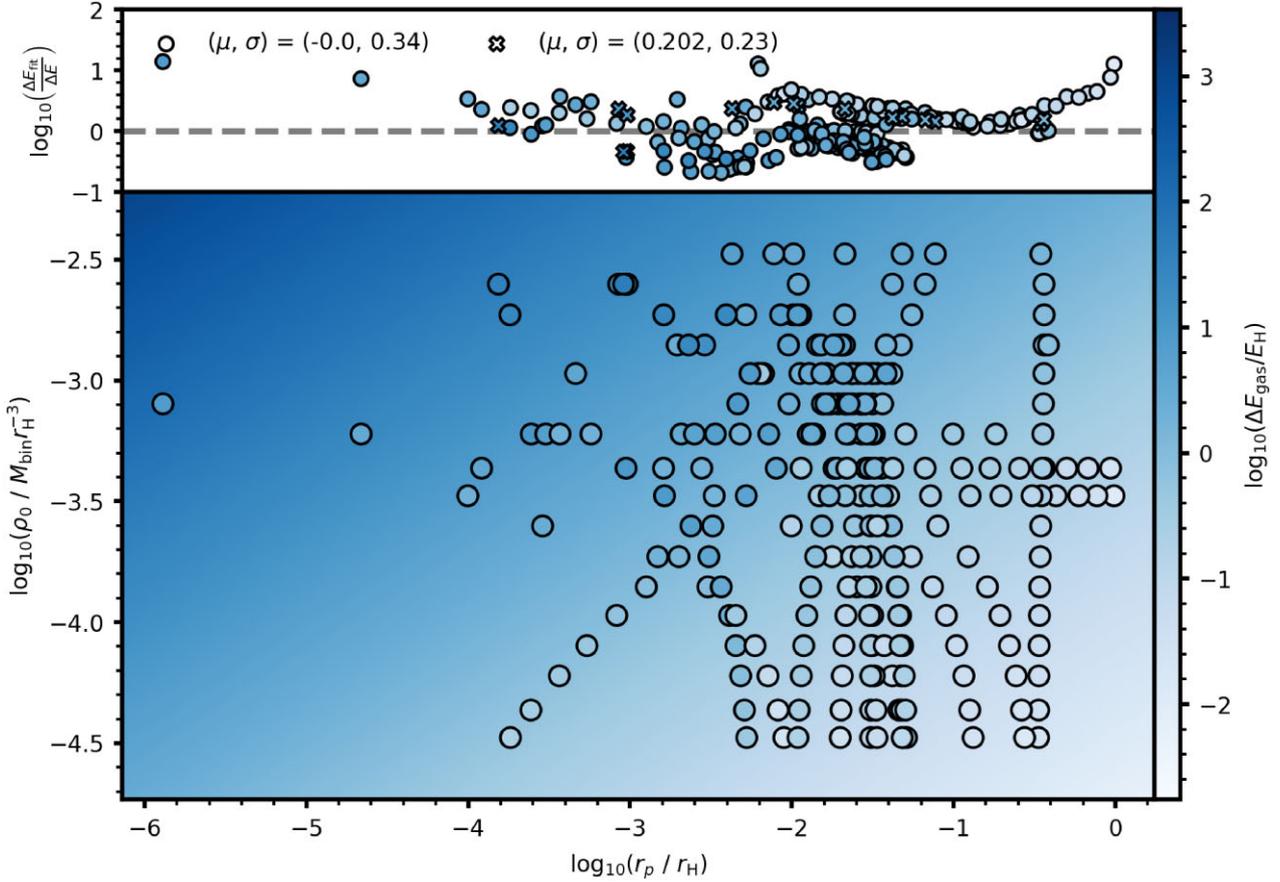

**Figure 12.** Energy changes due to gas dissipation during first encounter, fitted power laws in periapsis depth, and ambient disc density (see equation 24) for all 259 models featuring a close encounter. In the main plot, scatter markers are coloured by the true $\Delta E_{gas}$ values recorded by full hydrodynamical simulation. In the background, the mesh is coloured by the predicted $\Delta E_{gas}$ associated with that position in $\rho_0 - r_p$ space. Plotted above are the residuals, comparing the predicted dissipation to the true values. The residuals have an average scatter of 0.3 dex, not including those associated with the highest density runs with $\log_{10}\left(\frac{\rho_0}{\rho_s}\right) > -2.8$, which are excluded from the fit but included in this figure with cross markers to display the overestimation of the gas dissipation at high density.

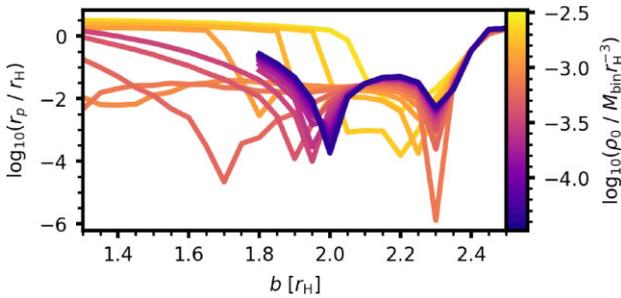

**Figure 13.** Periapsis depth at first close encounter as a function of initial impact parameter $b$, and ambient disc density $\rho_0$. At low densities the double valley structure matches that of the gas-less trajectories, with maximal angular momentum extracted by the SMBH for $b \sim 2.0$, $2.3 r_{\rm H}$. As density increases, the inner valley at $b = 2.0 r_{\rm H}$ starts to move inwards: at intermediate densities the centre of this valley can be found as low as at $b = 1.7 r_{\rm H}$. The outer valley at $b = 2.3 r_{\rm H}$ deepens with increasing density, then starts to shift inwards. Past $\rho_0 \sim 2.5 \times 10^{-3} M_{\rm bin} r_{\rm H}^{-3}$, these trends breakdown and we see a shift towards a single shallow, wide valley structure centred around $b = 2.2 r_{\rm H}$.

## 5.5 High density limit

Many of the rough correlations that hold at low density break down in the high density limit. This is, in part, due to the significant effect gas gravitation can have before close encounter. Within Fig. 10, we can identify failed binary formations for $b \leq 2.05 r_{\rm H}$ at high densities (lower right of panel). These outcomes arise due to self-interaction between a BH and its minidisc which result in it being perturbed away from the other BH before a close encounter can occur. By avoiding a close encounter and minidisc collision, the BHs do not undergo significant gas dissipation and so no binary is formed. The rapidity of this migration indicates that this effect is potentially non-physical. If the BH minidisc is not given sufficient time to stabilize before the BH is allowed to feel the gravity of the gas, irregularities in the disc may lead to a net acceleration not present in the equilibrium minidisc. This effect would likely be suppressed in hotter, more viscous discs as much of the minidisc substructure would be smoothed sooner. The effect is most pronounced for small impact parameters, as these systems take the longest time to reach close encounter and so the gas has more time to deflect the BHs from their equilibrium trajectories. Future studies may avoid this issue by initializing the BHs in a gas field closer to the steady minidisc state (i.e. improving on the methodology of Appendix C), or preventing the BHs from experiencing gas gravitation feedback until the BHs are within a certain distance.

Conversely, this pre-encounter deflection can actually lead to more binary formation at intermediate densities. Around $\log_{10}\left(\frac{\rho_0}{\rho_*}\right) \sim -3$ we observe an exceptionally wide area of successful capture. This capture area extends well below the $b \sim 1.6625 r_{\mathrm{H}}$ limit where close encounters stopped occurring in the gas-less case. Here, close encounters are still able to occur due to the pre-encounter gas-induced drift, which prevents the binary from performing the horseshoe orbits expected at such small radial separations. For a small range of densities near $\log_{10}\left(\frac{\rho_0}{\rho_*}\right) \sim -3$, this area of capture extends well beneath the main sampling grid; testing at a coarser $b$ resolution revealed that horseshoe orbits out-competed the effects of gas drift beneath $b \sim 0.5 r_{\mathrm{H}}$, resulting in no further close encounters/binary formations. As density is increased beyond $\log_{10}\left(\frac{\rho_0}{\rho_*}\right) \sim -3$ for small $b$ values, the strength of the migration is such that close encounters are again prevented, forcing wide flybys.

For systems with densities $\log_{10}\left(\frac{\rho_0}{\rho_*}\right) > -3$, almost all binaries that features a mutual Hill sphere intersection result in binary formation. The strength of gas gravitational dissipation can lead to the formation of very hard binaries after only a few periapsis passages. These binaries may be under-resolved by the gas grid, even with AMR, as the binary semimajor axis can approach the local cell size after multiple orbits. At the same time, as the disc gas density increases, we start to enter the regime where the effects of gas self-gravity (which is neglected in this study) may be significant. We can see this in the late time chaotic behaviour of high density systems, where the effect of gas gravitation can strongly perturb the binary orbits on short time-scales. For both of these reasons, we do not probe the parameter grid at higher disc densities. Nevertheless, we expect binary formation to be possible, and perhaps even more likely, in the high density regime, but the model in its current state is not designed for such a study.

Many interesting results can still be discerned from the higher density simulations. At large impact parameters, we observe captures without the need for deep close encounters. Fig. 14 details the evolution of a run with $b = 2.4 r_{\mathrm{H}}$ and $\rho_0 = 1.9 \times 10^{-3} M_{\mathrm{bin}} r_{\mathrm{H}}^{-3}$, where a prolonged dissipation event occurs at a relatively distant separation of $r_{\mathrm{p}} \sim 0.5 r_{\mathrm{H}}$ at $t = 41$ yr. The resulting binary is soft and retrograde after this first dissipation, but undergoes a more impulsive dissipation event at $t = 51$ yr during a very close second encounter. During the early binary evolution, the binary is unable to form a cohesive circumbinary disc, maintaining two spiral arms from the initial minidisc. This structure is disrupted by chaotic gas outflows during subsequent hardening periods.

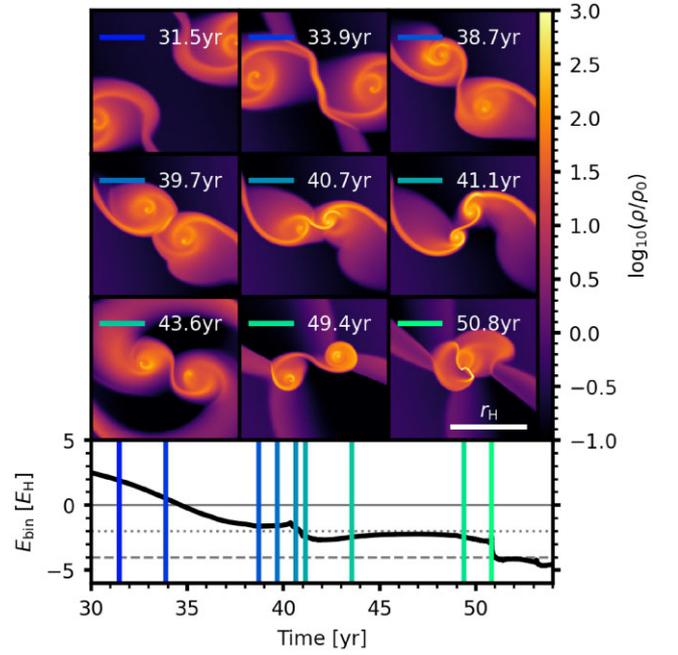

**Figure 14.** Prograde binary formation occurring within dense gas ($\rho_0 = 1.9 \times 10^{-3} M_{\mathrm{bin}} r_{\mathrm{H}}^{-3}$), widely separated ($b = 2.4 r_{\mathrm{H}}$) run, displaying the ability for capture to occur without severe close encounters in the presence of significant gas mass. The dissipation period at $t \sim 41$ yr is more extended than observed in the case of very deep close encounters, but a stable binary is formed regardless.

## 5.6 Orbital direction and inversion

A different way to visualize the outcome space is to consider the orbital direction of the binary close encounters, and their subsequent angular momentum evolution. Three families of orbits PI, RI, PE (prograde interior, retrograde interior, and prograde exterior, see Section 5.1) are expected even in the no-gas system due to the torque exerted by the SMBH. These families are distinct from the three regions A, B, C in which Jacobi encounters are predicted (see fig. 9 in Boekholt et al. 2023a). While the families PI, RI, and PE are distinguished by the orbital direction at first periapsis, the regions A, B, and C are defined by ranges in $b$ where multiple encounters between unbound BHs are driven by the SMBH. These multiple encounters are Jacobi encounters, and within each Jacobi region there are further subregions, creating a fractal hierarchy in $b$. While these regions were defined originally for the gas-less case, these multiple encounters can still be observed in low-density unbound (and weakly bound) systems.

Fig. 15 displays the full outcome grid with extra orbital information. In the low density limit, we identify the transition between the orbital families PI, RI, and PE: starting at small $b$ and increasing, we transition from prograde, to retrograde, back to prograde. As density is increased, the gas begins to perturb the BHs from their gas-less trajectories and the positions of these family boundaries shift inwards. The family boundaries represent zero-angular-momentum surfaces, with $L_{\mathrm{bin}}$ at periapsis ($L_{\mathrm{bin}}\big|_{r=r_{\mathrm{p}}} \equiv L(r_{\mathrm{p}})$) changing sign on either side. Because the binary flybys are approximately parabolic, systems near this surface also have the deepest first close encounters. Of the 133 binaries that form, 99 are initially retrograde (74 per cent). These binaries all form during retrograde interior (RI) encounters on intermediary impact parameters, representing the majority of formed binaries. In the high density limit however, we observe preferential

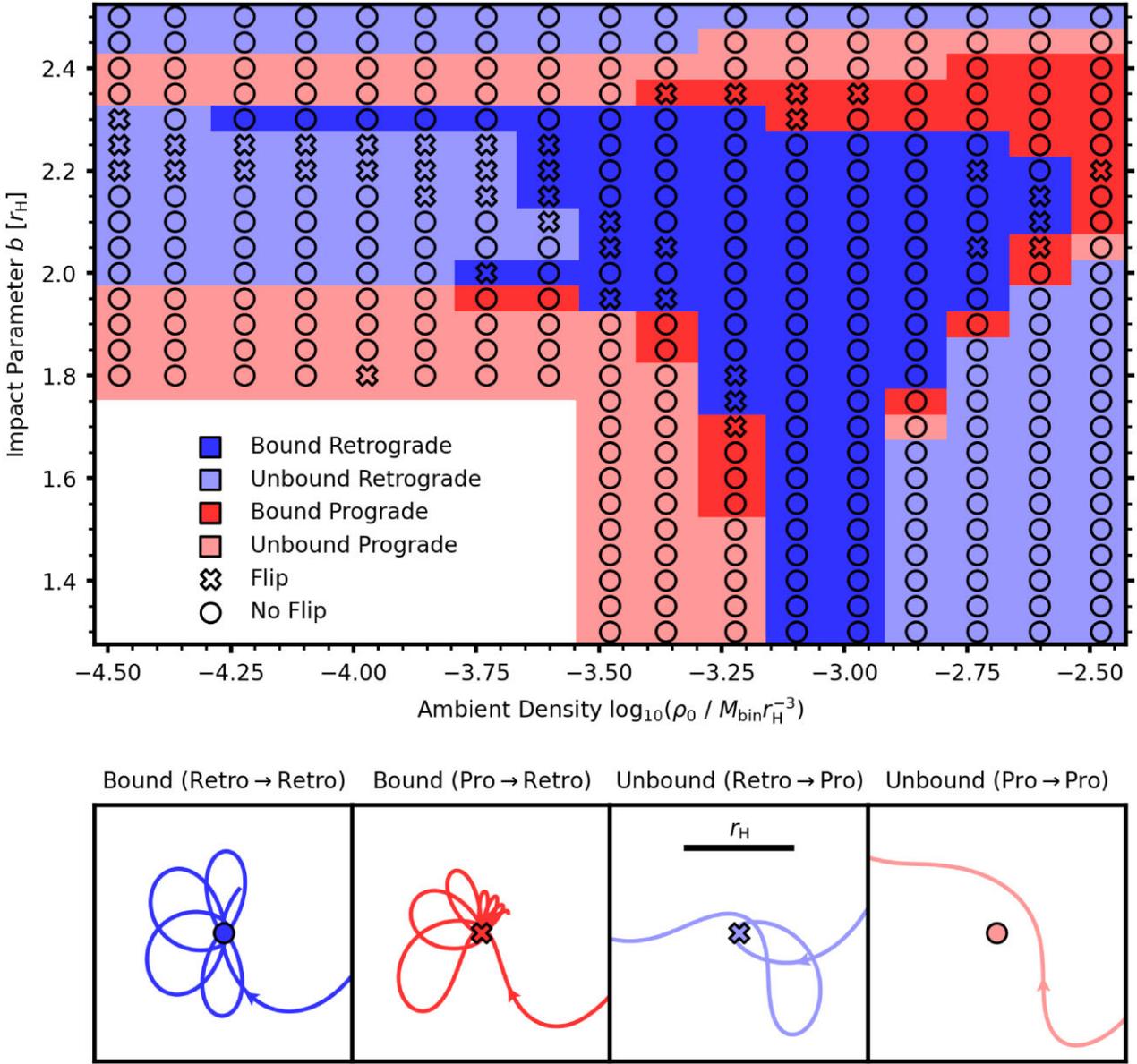

**Figure 15.** Outcome grid as in Fig. 10, but with sub-regions to better distinguish class of first close encounter (retrograde or prograde). For each simulation, marker type indicates the existence of orbit reversal in the post-encounter system. Multiple clear borders are found between prograde/retrograde encounters, along which orbit reversals may occur. Orbit reversal is also possible for the marginally bound retrograde systems near $\log_{10}\left(\frac{\rho_0}{\rho_*}\right) \sim -3.5$, where the SMBH torque remains significant post binary formation. The lower panel exhibits example encounters from the grid, with the trajectory of the outer BH traced in a frame centred on the inner BH.

prograde binary formation which may extend beyond our parameter grid.

Orbital inversion occurs when the binary angular momentum changes sign, resulting in a flip from prograde to retrograde or vice versa. In principle, this can be induced either by gas gravitation torques or by the SMBH but in our systems the latter tends to dominate. We inspect the grid for the presence of orbit reversal in the post-first-encounter evolution, which we identify by the existence of a later periapsis encounter within the Hill sphere with inverted angular momentum to the initial periapsis. Systems that experience an orbital inversion are marked as a 'flip' in Fig. 15.

Inversions are observed in both bound and unbound systems. For an unbound system to experience an inversion, they necessarily require a second close encounter. As such the unbound inversions that we record in Fig. 15 are a sign that we are observing Jacobi encounters. The large number of unbound inversions observed at low density between $b = 2.05 - 2.3 r_H$ are Jacobi encounters from Region B, made more easily visible as the light green region of Fig. G1 where systems are distinguished by their number of encounters. We also identify a few encounters from Region A near $b \sim 1.8 r_H$, though only one of these experiences an inversion.

Orbital inversion can also occur in bound systems. These inversions come in two types: Jacobi-like inversions and zero-angular-momentum inversions. In the former case, gas dissipation is significant enough to drive binary formation, but the young binary is still wide enough for the SMBH to drive the same inversion observed in the unbound case. These systems are observable in Fig. 15 at $\log_{10}\left(\frac{\rho}{\rho^*}\right) \sim -3.6$ between $b \sim 2.05 - 2.25 r_H$. They only perform a single inversion during their first period, such that the resulting stable binary is prograde despite the initial retrograde encounter. Orbital inversion is also possible near the zero-angular-momentum surfaces. Binaries formed near this surface have a very small absolute angular momentum after the first periapsis, and so very little torque is required to force them through an inversion. This is why even at higher densities, we see flips occurring near the prograde–retrograde boundaries.

Even in systems bound by gas dissipation, the actual inversion is consistently driven by SMBH torques. These torques are strongest when the binary is wide and soft, so we generally observe inversion within the first binary period, before significant hardening can occur. This helps explain why we do not observe any inversions within the central capture region away from the zero-angular-momentum surfaces: these systems have relatively large absolute angular momentum, and are too hard for the SMBH torque to be strong enough to drive inversion. It should be noted that the Jacobi-like inversions and zero-angular-momentum inversions are not totally distinct, the upper zero-angular-momentum surface at $b \sim 2.3 r_H$ overlaps with Region B of the Jacobi encounters.

### 5.7 Gravitational wave dissipation

It is expected that if the formed binary hardens over time the BHs will eventually be close enough for GW dissipation to be significant. Rowan et al. (2023b) found that some binaries resulting from gas capture were sufficiently eccentric to drive merger in a single periapsis passage. While we do not evolve our formed binaries for a long time after capture, we might still consider the comparative power of gas hardening to hardening by GW emission during only the first close encounter. The two processes share a general chronology for elliptical binaries: strong dissipation at periapsis followed by quiescence when the two bodies are distant. General relativistic effects are not included in our simulation (in a post-Newtonian form or otherwise), but we can still make predictions for the energetic contributions of GWs were they included. We can compute the energy lost at periapsis by Peters (1964).

$$\Delta E_{GW} = -\frac{\pi}{120} \frac{G^{\frac{7}{2}}}{c^5} \frac{M_{bin}^{9/2}}{r_p^{\frac{7}{2}}} g(e) \tag{28}$$

$$g(e) = 96 + 292 e^2 + 37 e^4 . \tag{29}$$

To get GW dissipation of a magnitude significant enough to affect the binary motion, we invert equation (28) to find the periapsis required for $\Delta E_{GW} \sim 0.1 E_H$, in the extreme case that $e \sim 1$. We find that, this requires a periapsis depth less than $r_p \sim 5(Gc^{-2}M_{bin}/r_H)^{5/7} r_H \sim 10^{-5} r_H$. Of all simulations, we only find a single run with this depth at first close encounter, $(b, \rho_0) = (2.3 r_H, 7.9 \times 10^{-4} M_{bin} r_H^{-3})$, which has an initial periapsis depth of $r_p \sim 10^{-6} r_H$. In this case, the periapsis is deep enough that a binary would form by GW dissipation without need for gas effects. However, as binary formation was already predicted by gas effects for this model, the non-inclusion of post-Newtonian physics has no qualitative effect on the outcome.

For our equal mass system and considering the limit $e \sim 1$, we can consider the boundary periapsis depth at which the energies dissipated by GW exceed that by gas dissipation (combining equations 28 and 24).

$$\left(\frac{r_p}{r_H}\right)^{\beta + \frac{7}{2}} < \frac{85\pi}{192A} \frac{G^{\frac{7}{2}}}{c^5} \frac{M_{bin}^{\frac{13}{2}}}{r_H^{\frac{9}{2}}} \left(\frac{\rho_0}{M_{bin} r_H^{-3}}\right)^{-\alpha} \tag{30}$$

For any specified SMBH–BH–BH configuration, we can predict the threshold beneath which we expect GW dissipation to dominate over gas dissipation. Within our system, this reduces to

$$\frac{r_p}{r_H} \lesssim 10^{-13} \left(\frac{\rho_0}{M_{bin} r_H^{-3}}\right)^{-0.33} \tag{31}$$

The prefactor $10^{-13}$ indicates that gas dissipation will remain dominant over GW dissipation for all but the deepest of close encounters or incredibly low density systems. In performing this calculation we have been forced to make a likely flawed assumption, asserting that the fitting procedure introduced in Section 5.3 will extrapolate down to exceptionally deep close encounters. We have only a few runs with a first periapsis beneath $\sim 10^{-4} r_H$, and we might expect that for very deep encounters the dependence of gas dissipation on depth may level off. If this is true, then GW dissipation may become more important for slightly shallower encounters. Regardless, we expect such extremely deep encounters to represent a very small proportion of the total encounters, so it remains fair to conclude that for the vast majority of encounters gas dissipation will be the dominant mechanism for binary formation. Once a stable binary forms, GW dissipation facilitates the inspiral and merger of the binary on much longer time-scales (Peters 1964), but this is beyond the scope of this study.

## 6 PREDICTIVE MODELS

The simulations presented above represent a large data set of failed and successful binary formations. From this data set, we attempt to construct an empirical predictive model that may allow one to obtain the outcome of the encounter without having to simulate the entire encounter. This is intended to improve on the currently used models (Goldreich, Lithwick & Sari 2002; Tagawa et al. 2020a; DeLaurentiis et al. 2023; Rozner et al. 2023), which typically rely on adjustments of the Ostriker (1999) gaseous dynamical friction formula valid in an infinite homogeneous medium and contradict the inferred energy dissipation rate in our simulations (equation 26). As we devise this model using data from simulations within our restricted parameter space, we do not necessarily expect the resulting predictions to be applicable to all gas-assisted formation systems. None the less, such a model can act as a benchmark for future studies into more generalized systems. To form this model, we use the dynamic properties of the binary at Hill sphere intersection to predict the conditions at periapsis. As discussed in Section 5.3, if the periapsis depth and ambient disc density are known, a reasonable estimation can be made for the energy dissipated by gas. We apply this modelling to the 259 models featuring a 'close encounter' where the BHs approach within $r = r_H$, as otherwise capture failure is assured.[5]

### 6.1 Predicting periapsis quantities

Each of the first encounters are approximately parabolic at periapsis; this approximation is strongest for $r_p \ll r_H$. As such, we have analytical means to calculate the periapsis depth provided the specific angular momentum at periapsis $L(r_p)$ is known:

$$r_p = \frac{L(r_p)^2}{2 G M_{bin} \mu^2} \tag{32}$$

---

[5] We also exclude the run with the very high gas mass $(\rho_0, b) = (3.3 \times 10^{-3} M_{bin} r_H^{-3}, 2.05 r_H)$ despite technically reaching $r = r_H$, as it features a very shallow turn-around trajectory that does not obey the general trends.

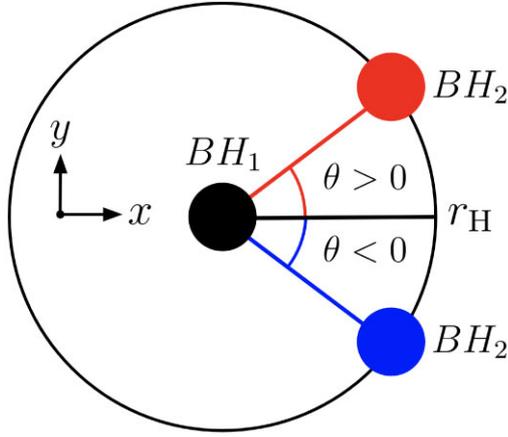

**Figure 16.** Diagram for the definition of the entry angle $\theta$. We measure $\theta$ when the two BHs first penetrate their mutual Hill sphere, between the SMBH radial direction (in the shearing frame $x$ direction) and the BH–BH separation vector. The angle is measured in the frame of the BH originating on the inner orbit.

We find that this relation holds approximately for all but the shallowest of first encounters (where $r_p \sim r_H$). The question then becomes how best to calculate the periapsis angular momenta. In the three-body SMBH–BH–BH system, $L_{bin}$ is not conserved and the torque exerted by the SMBH will depend on the BH trajectories. In comparing the binary angular momentum at Hill intersection ($L_{bin}|_{r=r_H} \equiv L(r_H)$) and at periapsis, we find a twin forked solution. Fig. 17 displays these two families of solutions and exhibits the separating factor: the entry angle into the Hill sphere $\theta$. Fig. 16 describes the geometry of $\theta$: the angle between the $x$ (or $R$) axis and the BH–BH separation upon first Hill sphere intersection. The amount of angular momentum exchanged between the binary and the SMBH varies depending on whether the outer BH leads or trails the inner BH on Hill intersection. As such, two branches of solutions form depending on the entry angle $\theta$. The trend is consistently linear outside of a complicated interaction area around $\theta \sim 0$; here the absolute angular momentum is large and so there is less need for a quality fit. In attempt to fit these branches, we apply three linear fits dependent on $L(r_H)$ and $\theta$, in the form $y = mx + c$, where $m$ and $c$ are specified in the legend in Fig. 17.

Angular momentum fitting is applied by first distinguishing the upper and lower angular momentum branches, and removing the elbow section at large negative $L(r_h)$. Any system with $\theta < 0$ necessarily belongs to the upper branch, and is used to fit the black dotted line. For systems with $\theta > 0$, we apply a cutoff at $L(r_H) = -0.8L_H$ due to the gradient change identified here. Systems with more negative angular momentum provide the fit for the red dashed line, the rest of the black dotted line. When predicting $L(r_p)$ for a new system, we first identify which branch it should belong to, then use the corresponding linear fit to identify its periapsis angular momentum.

With these two fits in hand, we can attempt to predict $r_p$ as a function of $L(r_H)$ and $\theta$. We find that the fit performs well for large $r_p$, but poorly for very deep encounters. This is not surprising, here small absolute changes in $L(r_p)$ can result in large changes in $r_p$ in log space. See Appendix F for a full breakdown of fitting errors. Propagating these values through, we now use the predicted $r_p$ values with the ambient gas density $\rho_0$ to calculate the energy dissipated by gas using equation (24). As shown in the bottom panel of Fig. F1, the majority of predicted values fall within 0.4 dex of the real values. The main contributors of error are due to miscalculations of periapsis depth for low-angular-momentum trajectories, and the relative large uncertainties generated from fitting equation (24).

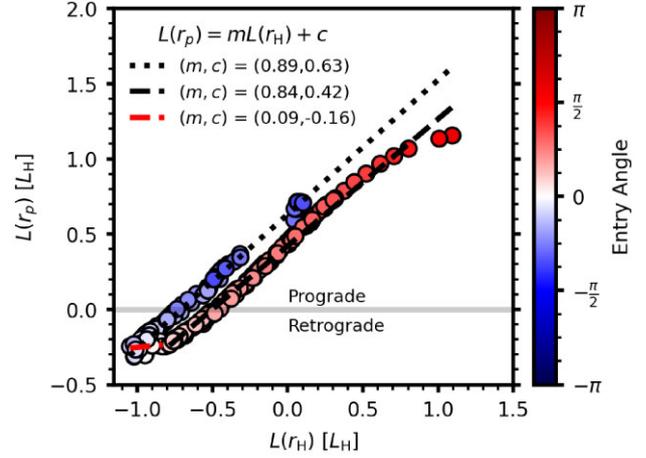

**Figure 17.** Twin forked structure in angular momentum space, when comparing angular momentum at Hill intersection $L(r_H)$, to angular momentum at periapsis $L(r_p)$. Membership to the two branches depends on the BH trajectory. If the outer BH leads the inner BH on Hill intersection, less angular momentum is injected. If the outer BH trails the inner BH on intersection, more angular momentum is injected. Within each family, a strong linear relationship is present, though this becomes more complicated near $\theta \sim 0$. The legend details fitting values for the three linear branches of form $y = mx + c$.

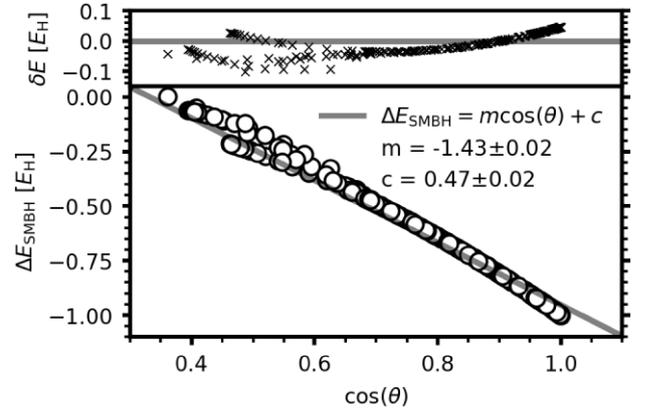

**Figure 18.** Correlation between entry angle $\theta$ with $\Delta E_{SMBH}$, the energy removed from the binary by the SMBH for all 259 simulations featuring mutual Hill sphere intersections. See Fig. 16 for details of the entry angle geometry. The relationship is well fit linearly in $\cos(\theta)$, though the residuals appears to be double valued at small $\cos(\theta)$ suggesting the dependence of $\Delta E_{SMBH}$ on some second unidentified parameter. We do not attempt to identify this parameter, as the fit remains of reasonably high quality regardless.

With a reasonable predictor for $\Delta E_{gas}$ formed, we now consider the energy dissipated by the SMBH. This proves to be a simple relation tied directly to the entry angle $\theta$, or more specifically the radial projection of BH separation at Hill sphere encounter $\cos(\theta)$ (see Fig. 18). This fit has very little spread from the true value, despite some unknown multivalued relationship at large negative $\theta$. SMBH dissipation is maximized when both of the BHs and the SMBH are in conjunction (small absolute $\theta$): here the SMBH tidal force acts directly against the BH relative velocity. For the more azimuthal aligned interactions (large absolute $\theta$), the SMBH tides are less aligned with velocity and so SMBH dissipation is lessened.

**Table 2.** Best-fitting parameter values and dispersion ($\sigma$) for the complete predictive modelling framework.

| Variable | Value | $\sigma$ |
|---|---|---|
| $\log_{10}(A)$ | 2.63 | 0.17 |
| $\alpha$ | 1.01 | 0.04 |
| $\beta$ | −0.43 | 0.03 |
| $B$ | −1.43 | 0.02 |
| $C$ | 0.47 | 0.02 |

We can now predict the binary energy immediately after the first periapsis passage as

$$E_{\rm bin,post}(r_{\rm p}, \theta, \rho_0) = E_{\rm bin}\big|_{r=r_{\rm H}} + \Delta E_{\rm gas}(r_{\rm p}, \rho_0) + \Delta E_{\rm SMBH}(\theta), \quad (33)$$

where $E_{\rm bin}\big|_{r=r_{\rm H}}$ is the initial energy of the BHs at Hill intersection, $\Delta E_{\rm gas}(r_{\rm p}, \rho_0)$ is given by equation (24) repeated here for convenience and $\Delta E_{\rm SMBH}(\theta)$ is the fit

$$\frac{\Delta E_{\rm gas}}{E_{\rm H}} = A\left(\frac{\rho_0}{M_{\rm bin}r_{\rm H}^{-3}}\right)^{\alpha}\left(\frac{r_{\rm p}}{r_{\rm H}}\right)^{\beta} \quad (34)$$

$$\frac{\Delta E_{\rm SMBH}}{E_{\rm H}} = B\cos(\theta) + C, \quad (35)$$

where $(A, \alpha, \beta, B, C)$ are fitting constants given by Table 2 and a periapsis depth $r_{\rm p}$ predicted by the parabolic approximation

$$r_{\rm p} = \frac{L(r_{\rm p})\mu^2}{2GM_{\rm bin}} \quad (36)$$

for a given periapsis angular momentum $L(r_{\rm p})$ predicted by $\theta$ and $L(r_{\rm H})$ the binary entry angle and angular momentum at Hill intersection

$$\frac{L(r_{\rm p})}{L_{\rm H}} = m\frac{L(r_{\rm H})}{L_{\rm H}} + c \quad (37)$$

$E_{\rm bin, post}$ acts as an predictive estimator of the expected binary energy immediately after the first periapsis. A complete breakdown of the errors introduced by each fitting step can be found in Appendix F. We use this energy estimator to predict the hardness of the binary and determine whether or not we expect the binary to remain bound.

### 6.2 Testing the predictive model

In determining the outcome of a close encounter we apply the same stability criterion introduced in Section 4.3: any binary which achieves $E_{\rm bin} < 2E_{\rm H}$ we expect to remain bound. We now use our predicted energies $E_{\rm bin, post}$ to assess what close encounters we expect to result in successful binary formation. Of the 345 models in the simulation grid, 259 feature a mutual Hill sphere intersection: the remaining 86 models are marked as unbound necessarily. We apply our predictive modelling to the 259 intersecting models, finding that our predictions match the true simulation outcome for 91 per cent of cases, as displayed in Fig. 19. Fig. 20 indicates that location of false positive and negatives, identifying that failed predictions are most likely on the fringe of the capture region identified in Fig. 10. We find our model is able to correctly predict the outcome of a flyby event for the vast majority of the simulations, despite the considerable scatter in exact predicted energy. This methodology may prove to be useful when coupled with global AGN simulations that attempt to predict BBH formation/merger rates without including a full hydrodynamical treatment, such as in Secunda et al. (2019) or Tagawa

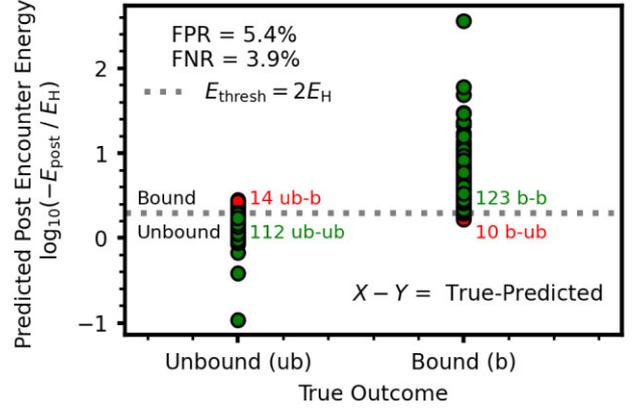

**Figure 19.** Accuracy of predictive modelling for the 259 simulations that feature a close BH–BH interaction. Prediction sets are labelled in format $X - Y$ where $X$ is the true outcome and $Y$ is the predicted outcome. While the predicted dissipated energy is sometimes very different from the true dissipation, the model succeeds in predicting the outcome the vast majority of the time: 91 per cent of models have the same predicted and 'true' outcome. The false positive and false negative rates (FPR and FNR, respectively) are only 5.4 per cent and 3.9 per cent.

et al. (2020a) which originally used simplified prescriptions for this process which are inconsistent with our numerical simulations. While our system only considers a subset of possible interaction types (no initial inclination, eccentricity), this is a step forward for semi-analytical models. Investigations into the effects of these variations would require a significant extension to the already expansive parameter space, this work is left for future studies. While these predictions may not generalize as well to formation environments outside our parameter space, its accuracy highlights the ability for the outcome of a complicated hydrodynamic system to be represented by relatively few dynamic/disc properties and sets the bar for future studies. We summarize each step of the predictive model below:

Summary of predictive binary capture model

Step I Identify system properties at Hill intersection: binary energy, angular momentum, entry angle, and ambient density

Step II Predict angular momentum at periapsis by the twin-fork fitting with equation (37) and Table 3

Step III Predict periapsis depth with the angular momentum at periapsis and equation (32)

Step IV Predict dissipation by gas via periapsis depth and ambient density with equation (34) and Table 2

Step V Predict dissipation by the SMBH via the entry angle with equation (35) and Table 2

Step VI Calculate the post-encounter energy $E_{\rm bin, post}$ by equation (33)

Step VII If $E_{\rm bin, post} < 2E_{\rm H}$, the binary is predicted to be bound after the first close encounter

## 7 DISCUSSION

### 7.1 Comparison to literature

There have been many recent papers that address the formation of BBHs in AGN discs. Their methodologies vary in terms of simulation type, domain, initial conditions, and simulated physics. In adopting initial conditions similar to that of CR1 we attempt to provide a clear comparison for gas-assisted BBH formation between SPH and a static Eulerian grid code (`PHANTOM` and `Athena++`). We explore a different density space to that of CR1; our most dense model is comparable to their low class mass system. As well as this, they also included accretion in their study, which we neglect, finding it to be the dominant driving force of binary dissipation during close encounters. Despite this, we report the same dissipation chronology: BHs form their own gaseous minidiscs which then collide during close encounters, dissipating orbital energy by gas gravitation. In both studies, binaries form successfully during only a single close encounter.

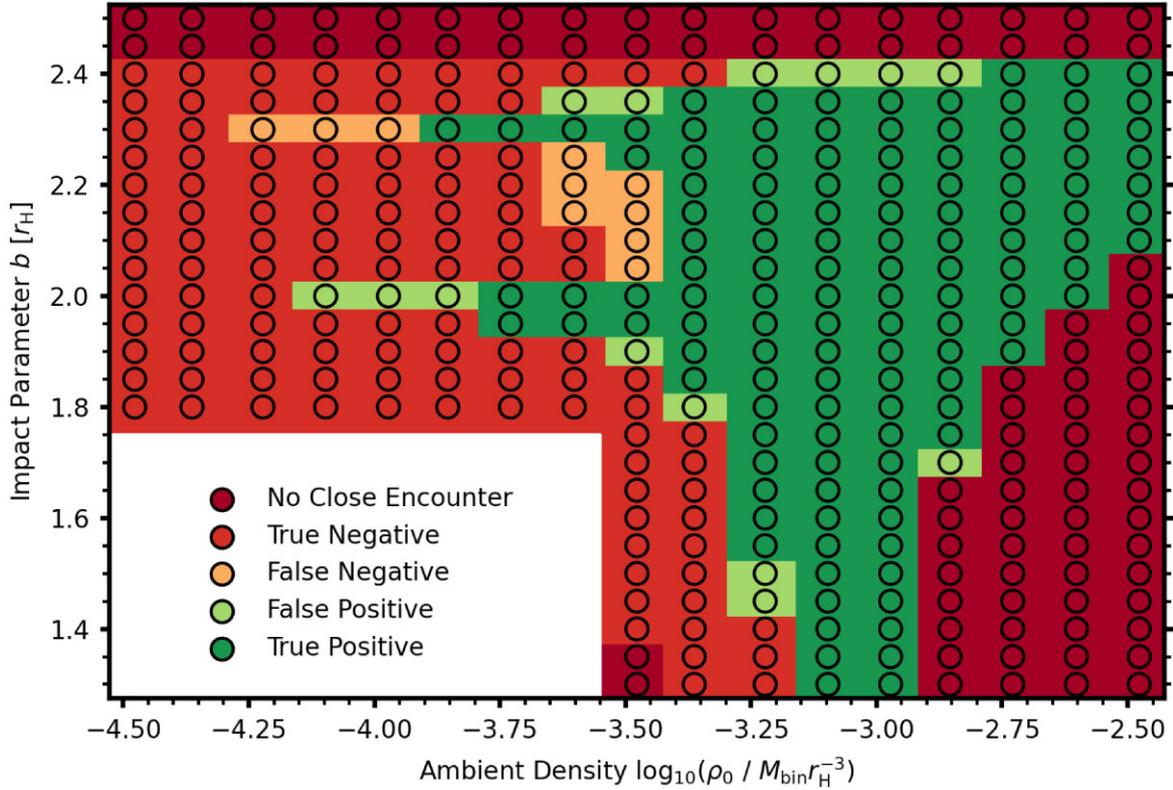

**Figure 20.** Predictor accuracy in the $b - \rho_0$ outcome space. Inaccurate predictions are located entirely on the fringe of successful binary formation, with the majority of the capture space well predicted. The model only makes incorrect predictions along the boundary between successful and unsuccessful capture, notably struggling to correctly predict binary formation for the very deep close encounters associated with $b = 2.0, 2.3 r_{\rm H}$. 86 models feature no close encounter and so necessarily do not form binaries (shown here in dark red).

**Table 3.** Best-fitting parameter values for angular momentum propagation from Hill sphere penetration to periapsis (see equation 37).

| m | c | Case |
|---|---|---|
| $0.89 \pm 0.008$ | $0.63 \pm 0.006$ | $\theta < 0$ |
| $0.84 \pm 0.008$ | $0.42 \pm 0.004$ | $\theta \geq 0$ and $L(r_{\rm H}) > -0.8L_{\rm H}$ |
| $0.09 \pm 0.098$ | $-0.16 \pm 0.083$ | $\theta \geq 0$ and $L(r_{\rm H}) \leq -0.8L_{\rm H}$ |

Li et al. (2023) analyses a similar capture scenario to our own, studying an annulus of the AGN gas disc using the 2D hydrodynamical grid code LA-COMPASS. They study a thicker, hotter disc with $H/R \sim 0.01$, and vary azimuthal BH–BH separation instead of radial, but they still observe BBH formation by gas gravitational dissipation alone, without the need for accretion. They only form retrograde BBHs, likely due to their selection of a single radial separation. They also put forward a predictive model for binary formation based on BBH energy at $r = 0.3r_{\rm H}$ and gas mass in the ambient Hill cylinder. We do not find significant agreement with this model, as our empirical fitting of gas dissipation with periapsis depth suggests that capture should still be possible for very low densities provided very deep close encounters are achieved. Moreover we do not find the BBH energy at $r = 0.3r_{\rm H}$ to be a significant determinant of outcome, as this energy has no clear correlation with periapsis depth (and subsequently the strength of gas dissipation). However, as with CR1, Li et al. (2023) considers a more massive gas density than our own study, equivalent to (and exceeding) our most massive models.

Recent semi-analytical studies (Tagawa et al. 2020a; Li, Lai & Rodet 2022; DeLaurentiis et al. 2023; Rozner et al. 2023) examined embedded binary formation by means of Ostriker (1999) gas dynamical friction (GDF). We do not find GDF to be an accurate description of the drag observed in our system, where the geometry of the gas distribution is especially important and so the assumption of a homogeneous gas background is inappropriate. Furthermore, we find that deeper and faster encounters dissipate more energy proportional to $v_{\rm p}$ as (equation 26) as opposed to in GDF where drag scales as $v_{\rm p}^{-2}$ with velocity (for supersonic trajectories). Despite this, these studies predict stable, eccentric BBH formations similar those observed in our hydrodynamic simulations. In particular, fig. 9 of DeLaurentiis et al. (2023) provides a similar $b - \rho_0$ cross-section to our own Fig. 10, displaying a similar extension of captures to smaller impact parameters as density increases as in our results, but that paper did not find captures at large $b$ when density is high. However, our simulations assumed initially corotating BHs with the AGN disc. It is possible that in systems with non-zero initial BH eccentricity or inclination, where each BH has a large relative velocity to the ambient gas, the formation of minidiscs would be suppressed. In this case, GDF may prove a more reliable prescription, but that case is beyond the scope of thisx paper.

The closest comparison can be made between this study and the companion study CR2, as it also considers interactions over a range of initial separations and two densities, albeit with a different numerical method, the smoothed particle hydrodynamics (SPH) code PHANTOM. The results are in broad agreement: we find encounters over a similar range of impact parameters, within the valley defined by encounters with the deepest periapsis. In both studies, the fractal flyby structure reported in the gas-less case by Boekholt et al. (2023a) is not observed, which is likely due to the relatively low resolution in the *b* space. The binaries in this paper lose similar amounts of energy by gas dissipation during close encounters as in CR2, generally on the scale of $1–10\,E_\mathrm{H}$. Most importantly, we observe very similar dependencies of gas dissipation on periapsis depth, where $\Delta E_\mathrm{gas} \propto r_\mathrm{p}^\beta$. In this work, we recovered $\beta = -0.43 \pm 0.03$ (equation 24 and Table 2), pleasingly similar to the $\beta = -0.42 \pm 0.16$ derived in CR2 though the latter features a considerably wider spread, perhaps influenced by the lower number of models available to fit with. The ability of studies with fundamentally different simulation methodologies (Eulerian AMR versus SPH) to reproduce this exponent shows promising indications that this is may be a intrinsic property of gas dissipation. It should be noted that the two studies are not in total agreement, CR2 implemented normalized initial azimuthal separations that kept the time to encounter between runs consistent. This likely explains why CR2 continues to observe a very wide capture window at high density, at which point this study recorded a reduction in cross section due to migratory effects (see Section 5.5). Moreover, CR2 implements a different predictive methodology for binary outcome, where dissipation by SMBH and gas effects are calculated together instead of separated. This model is likely to perform better in higher density systems where gas effects dominate the SMBH contribution. Outcome comparisons between the two models are hard to formulate, as the predictive model of CR2 utilizes empirically fit coefficients that are disc mass dependent, and only available for systems of higher disc density than included in this paper. They approach the system in a fundamentally similar way, utilizing the impact parameter at Hill intersection $p_\mathrm{1H}$ as a proxy for the true periapsis depth $r_\mathrm{p}$, which then determines the outcome of a potential formation event. The two methods complement each other by using generally similar parameters to predict capture for different regions of density space: lower densities at the fringe of capture in this study, higher densities more pertinent to physical disc systems in CR2.

There appears to be general agreement with both analytical and hydrodynamical models that close encounters between BHs in AGN discs should lead to the formation of highly eccentric, stable BBHs. We believe our wider parameter space, with fully hydrodynamical implementation, offers the best coverage of possible formation encounters presented in the literature so far. The empirical derivation of gas dissipation dependence on periapsis depth and gas density provides new insight into the physics of flyby hydrodynamics and prompts questions as to the validity of the GDF models commonly used to study these systems. The predictive modelling framework presents a clear methodology to better harmonize semi-analytical studies with full hydrodynamical simulations.

## 7.2 Limitations and caveats

In our simulations, we adopt several assumptions that should be considered when interpreting the results and may warrant further study:

(i) *BH feedback*: we have neglected any feedback from the BH into the AGN disc by jets or winds. Future studies may want to include a sub-grid prescription to account for these effects, even if the near-BH effects are not fully resolved.

(ii) *Gas self-gravity*: we have neglected gas self-gravity due to the increased computational expenditure required for such simulations. While this is a good approximation for the low ambient density models, as we increase the disc density this assumption weakens. Studies into higher disc masses should be careful in considering gas self-gravity as its inclusion will likely change gas morphology and subsequently the binary evolution.

(iii) *Thermodynamics*: we have adopted an isothermal equation of state to simplify our system. In real AGN systems, we expect contributions to the fluid energy and pressure from radiation, as well as heating and cooling effects. The competition between viscous dissipation, shock heating, and radiative cooling is likely to lead to variations in pressure throughout the gas, which may change minidisc viscosities and densities and subsequently the capture likelihood.

(iv) *Non-coplanar Interactions*: by restricting ourselves to a 2D simulation, we have considered only interactions that occur in the plane of the disc. We expect BHs in AGN to exists in orbits of varying inclinations, and previous studies have suggested that these inclinations should have a strong effect on capture/close encounter probability (Samsing et al. 2022).

(v) *Eccentricity*: we launch our black holes on Keplerian trajectories in the shearing box, simulating circular orbits in the AGN disc. Introducing eccentricity to the initial conditions will likely change the nature/probability of close encounters and non-zero BH velocity relative to the ambient gas may affect the ability for BHs to form stable mini-discs before encounter. Previous studies of eccentric and inclined interactions within AGN discs have already considered the gas-free case (Trani, Quaini & Colpi 2023).

(vi) *Shearing box*: we have simulated only a patch of a globally varying disc. In doing so we have massively reduced the computational expense, but are unable to consider the global effects such as potential gap clearing by the BHs. While our findings are consistent with CR1, CR2 which simulated a global disc annulus, this simplification may become more important as more massive interacting BHs are considered.

(vii) *Predictive modelling*: while our predictive model accurately describes our own systems, the strong connections between Hill quantities and periapsis quantities may break down when considering a wider spread of initial conditions e.g. BH eccentricity and inclination.

(viii) *Unresolved shocks*: despite use of AMR to finely resolve the gas morphology near the BH, we still experience some numerical difficulties in properly resolving the shock structure during close encounter. A more detailed discussion of the difficulties presented and solution adopted can be found in Appendix D.

(ix) *Synchronization*: the nature of the shearing box means models with smaller impact parameters take longer to reach close encounter. This means that they have longer to accumulate minidisc gas, a disparity between models. It can also lead to deflection from the equilibrium trajectories long before close encounter, which can either prevent or encourage close encounters. This study could be improved by offsetting the BH azimuthal separations to ensure similar encounter times for different impact parameters (such as implemented in CR2), although the agreement between the two studies suggest that the results are not sensitive to this assumption.

## 8 SUMMARY AND CONCLUSIONS

In this study, we simulated 345 flyby interactions between two isolated BHs on co-planar circular orbits in an AGN disc by spanning a parameter space in disc density and interactor impact parameter. We identified a region of feasibility where binary capture is expected to occur, accounting for 133 of the 259 systems featuring a close encounter. We introduced a predictive modelling framework for binary formation based on binary conditions at Hill sphere intersection. We summarize the key findings below:

(i) Dissipation by gas gravitation is sufficient to form stable binaries from isolated black holes in AGN discs. Upon approach, each BH perturbs the other's gaseous minidisc. When the discs collide, much of the minidisc gas is forced to trail behind the binary components, resulting in drag and dissipation of orbital energy. For close encounters, this dissipation occurs rapidly over a short period localized to the periapsis. Extended dissipation may also drive formation for wider encounters in more gas dense systems. We do not find the dissipative dynamics to be in good agreement with the gas dynamical friction prescriptions often applied to these systems.

(ii) We identify an extensive region in the impact parameter – initial ambient density ($b - \rho_0$) space where binary capture is efficient (Fig. 10). At low density, successful captures are restricted only to $b$ values where the close encounters are very deep, but the region widens as density increases. We resolve the transition of multiple separated capture islands at low density to a single large island at higher densities.

(iii) Capture likelihood is strongly tied to depth of first periapsis and only binaries that dissipate a significant proportion of their energy ($\Delta E \sim 2E_h$) during the first encounter form stable binaries. We find that the magnitude of energy dissipation by gas is well fit by power laws in periapsis depth and initial disc density such that

$$\Delta E_{gas} \propto \rho_0^{1.01 \pm 0.04} r_p^{-0.43 \pm 0.03} \tag{38}$$

This fit weakens at higher initial ambient densities, overestimating dissipation above $\rho_0 \sim 2.5 \times 10^{-3} M_{bin} r_H^{-3}$.

(iv) The measured dissipation of BH energy by gas is also roughly consistent with the simple expression

$$\Delta E_{gas} = 4.3 M_d v_H v_p, \tag{39}$$

where $v_H = (GM_{bin}/r_H)^{1/2}$ and $v_p = (2GM_{bin}/r_p)^{1/2}$ are the BHs' relative velocity for a parabolic orbit at separations of $2r_H$ and $r_p$, respectively, and $M_d$ is the mass of a minidisc at BH separations of $2r_H$.

(v) Gas assisted capture is possible even for very low initial ambient gas densities or minidisc mass, provided the initial impact parameter lies within the narrow valleys resulting in very close encounters (near the zero-angular momentum trajectories). In these encounters, the very high relative BH velocity as well as the strong disruption of each BH minidisc drives strong dissipation.

(vi) The binaries formed can be retrograde or prograde with respect to the AGN disc; while the former is more likely at lower densities, at higher densities the latter dominates (Fig. 15).

(vii) Formed binaries tend to be highly eccentric ($e > 0.8$), we do not observe a general tend towards circularization during the first $O(10)$ binary periods. The angular momentum evolution post formation depends on the hardness of the binary, SMBH effects tend to dominate unless the binary is very hard. Stable binaries may undergo orbital inversion due to SMBH effects within the first few periods.

(viii) The predicted dissipation by GW emission remains subdominant to gas dissipation for the vast majority of runs. We expect GW dissipation to dominate only for exceptionally deep close encounters at very low gas density. Gas effects dominate binary formation for a much wider range of impact parameters.

(ix) We provide accurate predictions as to the outcome of a binary formation event based only on quantities measurable at Hill sphere intersection (Section 6). For our system, these predictions match the hydrodynamically derived outcome 91 per cent of the time (Fig. 20). This modelling framework will be beneficial to semi-analytical studies by providing an inexpensive predictive methodology that is in strong agreement with the full hydrodynamical simulations while only using quantities available to $N$-body codes.

In simulating the hydrodynamics and $N$-body dynamics of the interaction between BHs in a gaseous disc, we affirm the ability for gas gravitation to encourage the formation of BBH in AGN discs and conduct a thorough analysis of the underlying physics processes.

## ACKNOWLEDGEMENTS

All simulations presented in this paper were performed on the Hydra cluster at University of Oxford. This work was supported by the Science and Technology Facilities Council Grant Number ST/W000301/1. TB's research was supported by an appointment to the NASA Postdoctoral Program at the NASA Ames Research Center, administered by Oak Ridge Associated Universities under contract with NASA.

## DATA AVAILABILITY

The data underlying this article will be shared on reasonable request to the corresponding author.

## APPENDIX A: CONSTANT VISCOSITY

In this study, we adopt a homogeneous viscosity throughout the simulation domain. In real system, we expect viscosity to vary throughout the shearing box due to variations in the local sound speed $c_s$ and compression of the minidiscs vertically by the gravity of each BH. Li & Lai (2023a) adopt a reasonable prescription within their paper by prescribing a variable $\Omega$ within the expression $\nu = \frac{c_s^2}{\Omega}$ that considers BH contributions within the SMBH gravity field. However, we feel that without a full treatment with more realistic heating and cooling physics, accurate predictions as to the thickness of the BH minidiscs and therefore the size of turbulent vortices are impossible to make. Such physics are beyond the scope of this study and we instead adopt a homogeneous viscosity as a first-order approximation to a system that requires further advanced modelling to more accurately simulate.

## APPENDIX B: TEMPORAL CONVERGENCE

Careful study of our system at very low gas density and for very deep close encounters made clear the need for caution when attempting to resolve the motion of the BHs temporally. If an insufficiently small pre-factor $\eta$ is chosen to scale the BH time-step (as introduced in Section 2.2), an error arises in the $n$-body integration during close encounter. Here, where the calculated time-step is too large, the binary system loses energy spuriously due to numeric effects, leading to the creation of a binary that should not exist physically. This pseudo-dissipation is most noticeable for systems with a very deep first periapsis. Fig. B1 displays the energy error after first close encounter for runs with larger $\eta$ values, as compared to the $\eta = 0.001$ system. Investigation into convergence for such systems suggests that for this study, $\eta \leq 0.005$ is sufficient, though we set $\eta = 0.001$ for safety. For a more comprehensive study covering much deeper close encounters, a more stable integrator or smaller $\eta$ value may be required.

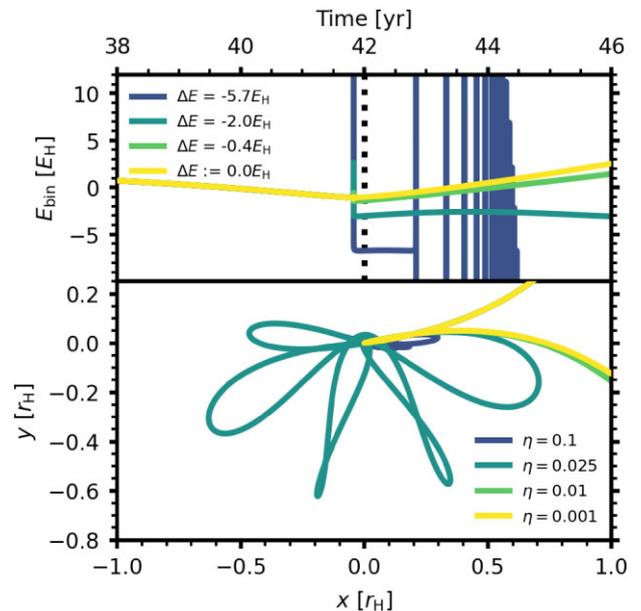

**Figure B1.** Comparisons of runs with $b = 2.0r_H$, $\rho_0 = 0$ with varying values for the time-scale pre-factor $\eta$. In the lower panel, the path traced by the outer BH in a frame centred on the inner BH. In the top panel, the binary energy as a function of time including errors in energy. If $\eta$ is too large, a significant energy error is generated during first close encounter. In the cases where $\eta \geq 0.025$, the error is so severe that a binary is spuriously formed. If $\eta$ is reduced, this issue is not present. Energy errors $\Delta E$ are calculated in comparison to the $\eta = 0.001$ model at $t = 42$ yr (marked by the dotted vertical line in the top plot).

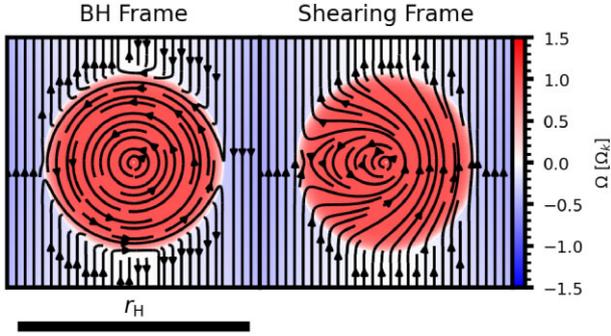

**Figure C1.** Velocity field near the inner BH at $t = 0$, displayed with streamlines in the BH frame (left) and shearing frame (right). The background colourmap displays orbital frequency about the BH in its own frame.

### APPENDIX C: SEEDING BLACK HOLES

As introduced in Section 3, our BHs are added impulsively to the disc at $t = 0$, as opposed to growing the BH mass steadily from zero. Without care, this can lead to expensive-to-resolve chaotic flows as gas falls directly onto each BH. To stabilize this early phase, we seed each BH within a distorted velocity field. We do not find this seeding to effect the late-time behaviour or outcomes and it is included mostly as a precaution. Under this routine, the true gas velocity field $\boldsymbol{u}$ is dependent on proximity to the BHs.

$$\boldsymbol{u} = \begin{cases} \boldsymbol{u}_0 & r_n \geq r_d \forall n \\ \boldsymbol{u}_0 (1 - k_d) + k_d \boldsymbol{u}_{d,n} & r_n < r_d \end{cases}, \tag{C1}$$

where we interpolate between velocity fields for the background $\boldsymbol{u}_0$ and for minidisc $n$ $\boldsymbol{u}_{d,n}$ by a kernel $k_d = 1 - \left(\frac{r_n}{r_d}\right)^{16}$ on a length scale $r_d = 0.5 r_{\mathrm{h, s}}$. Each of the velocity fields are represented as

$$\boldsymbol{u}_0 = \boldsymbol{u}_{\mathrm{eq}} \begin{pmatrix} 0 \\ -q\Omega x \end{pmatrix} \tag{C2}$$

$$\boldsymbol{u}_{d,n} = \sqrt{\frac{Gm_n}{r_n^2}} \hat{\boldsymbol{\phi}}_n + \boldsymbol{v}_n \tag{C3}$$

Here the background field is the Keplerian shear of the AGN disc and the local velocity field matches Keplerian rotation around an isolated, mobile BH labelled $n$. This interpolated velocity does not match the steady-state velocity field reached once the minidiscs have formed, but acts as a stabilizing intermediary solution with an easy analytic representation. Fig. C1 demonstrates this routine by displaying the velocity field near a BH at $t = 0$.

### APPENDIX D: NUMERICAL ISSUES

For a small subset of runs in the parameter grid, we experienced numerical errors arising from strong density gradients generated during the chaotic gas flows immediately after close BH encounter. This would lead to large non-physical accelerations of single cells of gas. Multiple methods were attempted to treat this issue, involving a variety of different solvers, reconstructions, resolutions, and integrators. It is possible that such instabilities could be resolved fully if the resolution were increased considerably near each BH, but this would lead to a computational cost unacceptable for this study. Ultimately, no one choice was able to eliminate 100 per cent of the errors. As such, we were forced to adopt an 'error-catching' routine that identified failing cells and smoothed out their hydrodynamical quantities with their neighbours. As only a single cell would fail in a single time-step, we believe this method prevented the arrival of such errors from affecting the large-scale behaviour of the hydrodynamical system. We do not expect the presence of such errors, or the corrective methodology implemented, to have any effect on the conclusions drawn by this study.

### APPENDIX E: MINIDISC MASS GROWTH

Under the isothermal equation of state, the fluid equations of motion (see equation 2) are linear in density. This means that minidiscs growing from systems with different ambient densities are morphologically similar but with proportionally different minidisc masses. Given that each simulation evolves for a similar amount of time before close encounter (should one occur), we can predict the minidisc mass at interaction from the ambient AGN disc density. Fig. E1 details the strong linear relationship between the ambient density and the gas mass within the Hill sphere of a single BH when the BHs reach within $2 r_{\mathrm{H}}$ of each other. There is only a minor variance between simulations at different radial separations, due to slightly different traveltimes to intersection. We note that some models at high density have considerably less massive minidiscs, this is due to gas perturbing the BHs onto wide, fast moving trajectories which prevent significant gas buildup. This is generally inconsequential to the analysis presented in this paper, as such trajectories feature no close encounter.

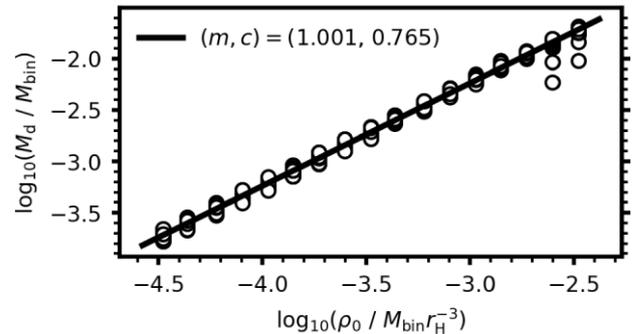

**Figure E1.** Linear relationship between the ambient AGN disc density and the approximate minidisc masses before close encounter for models with $b = 1.8, 2.5 r_{\mathrm{H}}$. There is some variation in disc mass between models with different $b$ values at set densities, and some of the higher density systems feature under-massive minidiscs, but generally the fit holds well.

## APPENDIX F: PREDICTIVE MODELLING ACCURACY

The framework presented in Section 6 provides a reasonably accurate (0.4 dex) predictor of binary energy immediately after first periapsis given the binary orbital components are known at Hill intersection. The model itself performs better in certain areas than others: it struggles to accurately predict very deep periapsis encounters, and the energetic fitting from equation (24) is a significant source of scatter. Fig. F1 gives a complete breakdown of the errors introduced by each step of the fitting procedure.

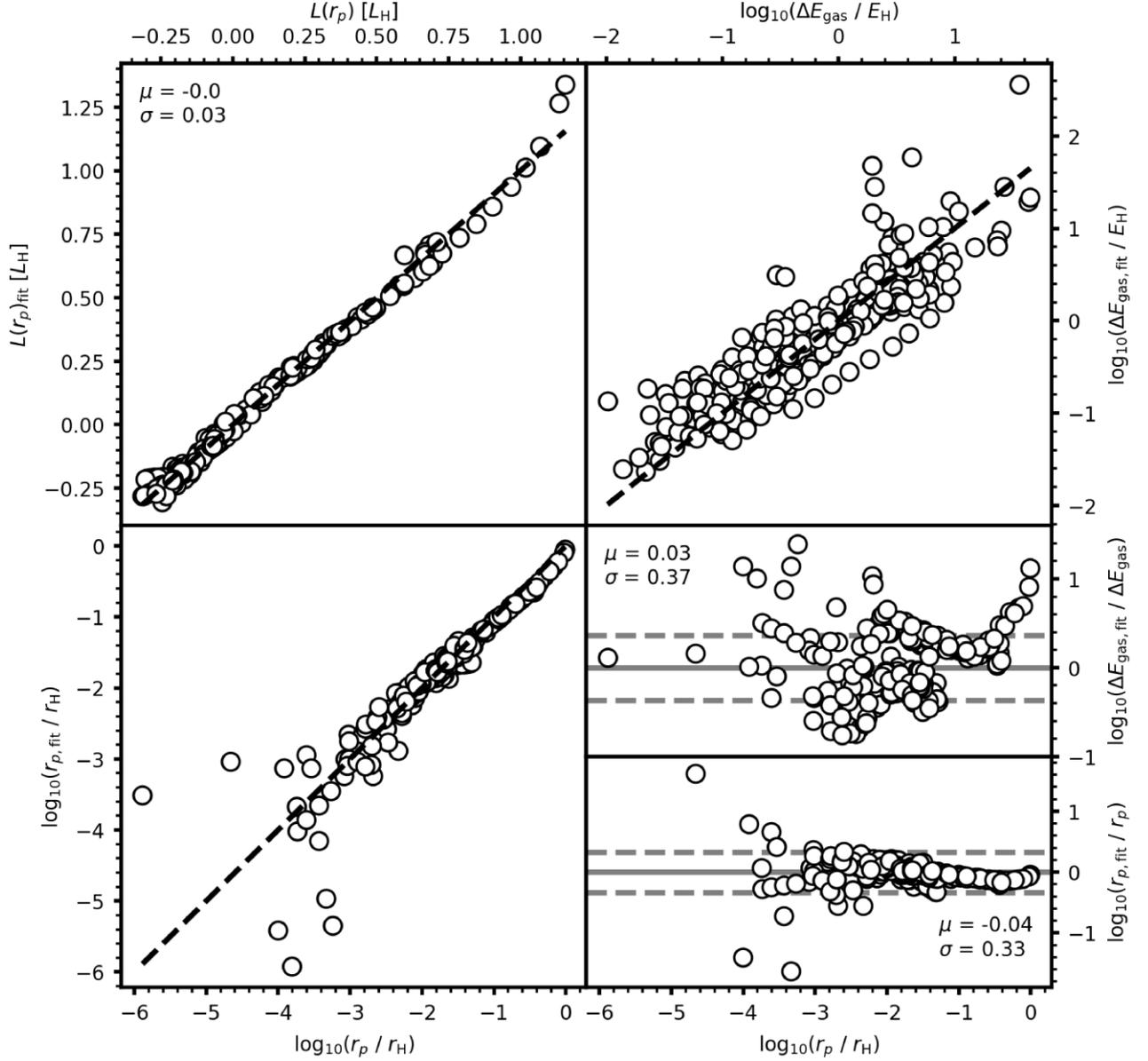

**Figure F1.** Fitting data for the predictive modelling sub-routines, for all 259 models that penetrate the mutual Hill volume. Top left panel: comparison of true angular momentum to that predicted by fitting from Fig. 17, where the legend details the mean and spread in the error. Bottom left panel: comparison of true periapsis depth to that predicted by periapsis momentum and equation (32). Top right panel: comparison of true gas dissipated energy to that predicted by full predictive modelling from the Hill quantities. Bottom right panels: residuals for upper right ($\Delta E_{gas}$) and lower left ($r_p$) panels, with $1\sigma$ bounds as dashed lines. Energetic predictions by Hill intersection propagated values hold on to on average 0.4 dex, with the majority of scatter introduced by the power-law fit from equation (24). The predictor consistently overestimates the dissipation for very shallow encounters, likely due to the inaccuracy of equation (32) for $r_p \sim r_H$. Note that these residuals are different from those presented in Fig. 12, as in that method $r_p$ was measured directly, whereas here it is predicted.

## APPENDIX G: JACOBI REGIONS

While not the principle focus of this paper, we can inspect our grid of simulations for the existence of Jacobi interactions: narrow regions in impact parameter space where the SMBH is able to encourage multiple interactions between the black holes even without the presence of gas (Boekholt et al. 2023a). These interactions are identifiable as unbound systems with multiple interactions i.e. systems with a second periapsis within the Hill sphere. Such systems are highlighted in Fig. G1.

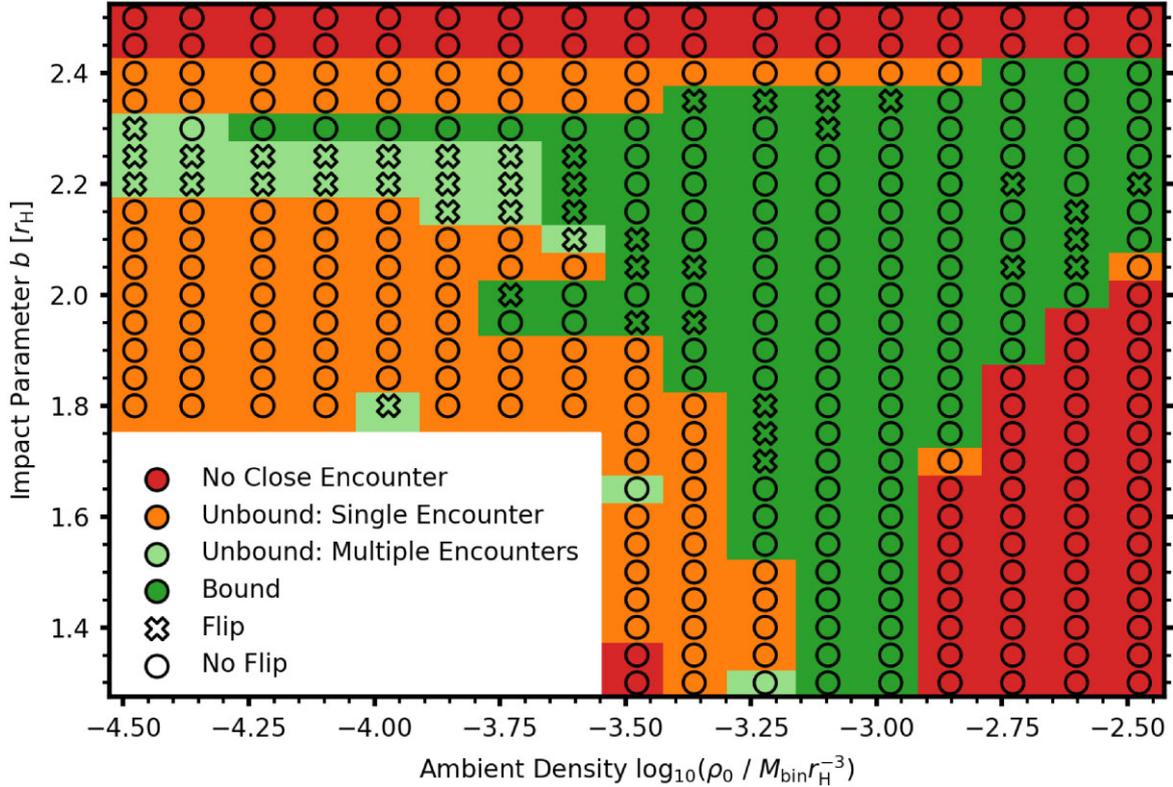

**Figure G1.** Number of close encounters between BHs for the full grid, where a close encounter is defined as a periapsis within the mutual Hill sphere. Jacobi encounters are visible around $b = 2.15 - 2.3 r_H$ in light green within Region B [as defined in Boekholt et al. (2023a)], but also in a couple of systems around $b = 1.6 - 1.8 r_H$ which are likely members of Region A. Region C is substantially narrower than the other regions, and likely lies beneath the sampling resolution in $b$ used in this study (we might expect to see it around $b = 2.45 r_H$). Inversion can also occur for bound systems near the zero angular momentum surface; these are not true Jacobi interactions (see Section 5.6).

This paper has been typeset from a TEX/LATEX file prepared by the author.